\documentclass[11pt]{article}
\usepackage{amsmath,amsthm,latexsym,amssymb,amsfonts,epsfig,subfigure, psfrag,url,dsfont}

\makeatletter \@addtoreset{equation}{section} \makeatother

\newtheorem{Theorem}{Theorem}[section]

\def\be{\begin{equation}}
\def\ee{\end{equation}}
\def\ba{\begin{eqnarray}}
\def\ea{\end{eqnarray}}

\newcommand\nn{\nonumber}
\newcommand\q{\quad}
\def\Nl{{\mathchoice
{\setbox0=\hbox{$\displaystyle\rm N$}\hbox{\hbox to0pt
{\kern0.4\wd0\vrule height0.9\ht0\hss}\box0}}
{\setbox0=\hbox{$\textstyle\rm N$}\hbox{\hbox to0pt
{\kern0.4\wd0\vrule height0.9\ht0\hss}\box0}}
{\setbox0=\hbox{$\scriptstyle\rm N$}\hbox{\hbox to0pt
{\kern0.4\wd0\vrule height0.9\ht0\hss}\box0}}
{\setbox0=\hbox{$\scriptscriptstyle\rm N$}\hbox{\hbox to0pt
{\kern0.4\wd0\vrule height0.9\ht0\hss}\box0}}}}
\def\Zl{{\mathchoice
{\setbox0=\hbox{$\displaystyle\rm Z$}\hbox{\hbox to0pt
{\kern0.4\wd0\vrule height0.9\ht0\hss}\box0}}
{\setbox0=\hbox{$\textstyle\rm Z$}\hbox{\hbox to0pt
{\kern0.4\wd0\vrule height0.9\ht0\hss}\box0}}
{\setbox0=\hbox{$\scriptstyle\rm Z$}\hbox{\hbox to0pt
{\kern0.4\wd0\vrule height0.9\ht0\hss}\box0}}
{\setbox0=\hbox{$\scriptscriptstyle\rm Z$}\hbox{\hbox to0pt
{\kern0.4\wd0\vrule height0.9\ht0\hss}\box0}}}}
\def\Ql{{\mathchoice
{\setbox0=\hbox{$\displaystyle\rm Q$}\hbox{\hbox to0pt
{\kern0.4\wd0\vrule height0.9\ht0\hss}\box0}}
{\setbox0=\hbox{$\textstyle\rm Q$}\hbox{\hbox to0pt
{\kern0.4\wd0\vrule height0.9\ht0\hss}\box0}}
{\setbox0=\hbox{$\scriptstyle\rm Q$}\hbox{\hbox to0pt
{\kern0.4\wd0\vrule height0.9\ht0\hss}\box0}}
{\setbox0=\hbox{$\scriptscriptstyle\rm Q$}\hbox{\hbox to0pt
{\kern0.4\wd0\vrule height0.9\ht0\hss}\box0}}}}
\def\Rl{{\mathchoice
{\setbox0=\hbox{$\displaystyle\rm R$}\hbox{\hbox to0pt
{\kern0.4\wd0\vrule height0.9\ht0\hss}\box0}}
{\setbox0=\hbox{$\textstyle\rm R$}\hbox{\hbox to0pt
{\kern0.4\wd0\vrule height0.9\ht0\hss}\box0}}
{\setbox0=\hbox{$\scriptstyle\rm R$}\hbox{\hbox to0pt
{\kern0.4\wd0\vrule height0.9\ht0\hss}\box0}}
{\setbox0=\hbox{$\scriptscriptstyle\rm R$}\hbox{\hbox to0pt
{\kern0.4\wd0\vrule height0.9\ht0\hss}\box0}}}}
\def\Cl{{\mathchoice
{\setbox0=\hbox{$\displaystyle\rm C$}\hbox{\hbox to0pt
{\kern0.4\wd0\vrule height0.9\ht0\hss}\box0}}
{\setbox0=\hbox{$\textstyle\rm C$}\hbox{\hbox to0pt
{\kern0.4\wd0\vrule height0.9\ht0\hss}\box0}}
{\setbox0=\hbox{$\scriptstyle\rm C$}\hbox{\hbox to0pt
{\kern0.4\wd0\vrule height0.9\ht0\hss}\box0}}
{\setbox0=\hbox{$\scriptscriptstyle\rm C$}\hbox{\hbox to0pt
{\kern0.4\wd0\vrule height0.9\ht0\hss}\box0}}}}
\def\Hl{{\mathchoice
{\setbox0=\hbox{$\displaystyle\rm H$}\hbox{\hbox to0pt
{\kern0.4\wd0\vrule height0.9\ht0\hss}\box0}}
{\setbox0=\hbox{$\textstyle\rm H$}\hbox{\hbox to0pt
{\kern0.4\wd0\vrule height0.9\ht0\hss}\box0}}
{\setbox0=\hbox{$\scriptstyle\rm H$}\hbox{\hbox to0pt
{\kern0.4\wd0\vrule height0.9\ht0\hss}\box0}}
{\setbox0=\hbox{$\scriptscriptstyle\rm H$}\hbox{\hbox to0pt
{\kern0.4\wd0\vrule height0.9\ht0\hss}\box0}}}}
\def\Ol{{\mathchoice
{\setbox0=\hbox{$\displaystyle\rm O$}\hbox{\hbox to0pt
{\kern0.4\wd0\vrule height0.9\ht0\hss}\box0}}
{\setbox0=\hbox{$\textstyle\rm O$}\hbox{\hbox to0pt
{\kern0.4\wd0\vrule height0.9\ht0\hss}\box0}}
{\setbox0=\hbox{$\scriptstyle\rm O$}\hbox{\hbox to0pt
{\kern0.4\wd0\vrule height0.9\ht0\hss}\box0}}
{\setbox0=\hbox{$\scriptscriptstyle\rm O$}\hbox{\hbox to0pt
{\kern0.4\wd0\vrule height0.9\ht0\hss}\box0}}}}
\newcommand{\cc}{\mathcal C}

\newcommand{\ch}{\mathcal H}

\newcommand{\cq}{\mathcal Q}

\def\nn{\nonumber}

\newcommand{\eqa}{\begin{eqnarray}}
\newcommand{\neqa}{\end{eqnarray}}

\usepackage{bbm}

\def\q{{\quad}}

\begin{document}



%

%

{\renewcommand{\thefootnote}{\fnsymbol{footnote}}
\hfill ITP--UU--11/42, SPIN--11/33, IGC--11/11--1\\\vspace{1.5cm}
\begin{center}
{\LARGE  Effective relational dynamics of a nonintegrable cosmological model}\\
\vspace{2em}
Philipp A H\"ohn\footnote{e-mail address: {\tt phoehn@perimeterinstitute.ca}}$^{1}$,
Em\'ilia Kubalov\'a\footnote{e-mail address: {\tt milka@physics.muni.cz}}$^{2}$, and
Artur Tsobanjan\footnote{e-mail address: {\tt
artur.tsobanjan@gmail.com}}$^3$
\\
\vspace{1em}
$^1$ Institute for Theoretical Physics,\\
 Universiteit Utrecht,\\
Leuvenlaan 4, NL-3584 CE Utrecht, The Netherlands

\vspace{1em}
$^2$ 
Department of Theoretical Physics and Astrophysics,\\
Faculty of Science, Masaryk University,\\
Kotl\'a\v{r}sk\'a 2, 611 37 Brno, Czech Republic\\

\vspace{1em}
$^3$Institute for Gravitation and the Cosmos,\\
The Pennsylvania State
University,\\
104 Davey Lab, University Park, PA 16802, USA\\
\vspace{1em}

\vspace{1.5em}
\end{center}
}


\setcounter{footnote}{0}

\begin{abstract}
We apply the effective approach to evaluating semiclassical
relational dynamics to the closed Friedman--Robertson--Walker
cosmological model filled with a minimally coupled massive scalar
field. This model is interesting for studying relational dynamics in
a more general setting because (i) it features a nontrivial
coupling of the relational clock to the evolving degrees of freedom,
(ii) no temporally global clock variable exists, and (iii) it is
nonintegrable which is typical for generic dynamical systems. The
effective approach is especially well geared for addressing the
concept of relational evolution in this context, since it enables one
to switch between different clocks and yields a consistent
(temporally) local time evolution with transient observables so long
as semiclassicality holds. We provide evidence that relational
evolution in this model universe, while possible for sufficiently
semiclassical states, generically breaks down in the region of
maximal expansion. This is rooted in a defocusing of classical
trajectories, which leads to a rapid spreading of states that are initially sharply peaked
and to a mixing of internal time directions in this region. These
results are qualitatively compared to previous work on this model,
revisiting conceptual issues that have been raised earlier in the
literature.
\end{abstract}


\section{Introduction}

How can a unitary evolution in a `classical' time emerge from the
full quantum theory? This question is one of the central
conundrums in quantum gravity and cosmology and constitutes one of
the many facets of the problem of time
\cite{pot,isham1,bht1,bht2,zeh}. It is rooted in the absence of a
time coordinate in the quantum theory and in the necessity to
instead employ dynamical degrees of freedom
to keep track of (an internal) time
\cite{rel,isham}. Such relational clock variables, however, are not
perfect monotonic and classical clocks whose increment coincides
with the increment of some observer's proper time. Rather, they are
genuine quantum degrees of freedom subject to quantum fluctuations
and even classically will generically not always run forward,
leading to what is
known as the {\it global problem of time}
\cite{pot,isham1,bht1,bht2}. Imperfect relational clocks generically
couple to other degrees of freedom of the system, which causes
backreaction and complicates a good resolution of the evolution of
the remaining degrees of freedom in such a clock
\cite{marolf1,marolf2,gidmarhar}. In particular, a good resolution
of unitary relational quantum evolution requires an approximate
division between the degrees of freedom to be measured and the
clock. This division is state dependent and may, in fact, become
impossible in highly quantum states
\cite{bht1,bht2,marolf1,marolf2,gidmarhar}. The challenge of
recovering a unitary evolution in a `classical' internal time is
thus a highly nontrivial one even in the semiclassical regime.

Extracting dynamical information from finite dimensional systems as
in quantum cosmology is generally achieved by
deparametrizations in specific matter degrees of freedom, such as
dust or free scalar fields (or model--specific geometrical degrees of
freedom \cite{bianchi}), which assume the role of internal clocks,
and a lot of progress has been made in this direction
\cite{martinbooks,lqc}. However, the standard free scalar field
\cite{lqc}, as well as the recently discussed dust fields
\cite{dust} decouple from the other degrees of freedom, yield a
`time--independent' Hamiltonian and correspond to the `ideal clock
limit' of \cite{marolf1}. These matter clocks are therefore rather
special in nature.

Furthermore, the issue of nonintegrability, despite being the generic case in dynamical systems \cite{chaos} and having severe implications for relational evolution, has largely been overlooked in the literature on relational dynamics. Specifically, in such a situation, the only global constant of motion (i.e.\ Dirac observable for constrained systems) is the Hamiltonian (constraint) \cite{chaos}. Nevertheless, (relational) observables can still exist implicitly and locally, and thus relational evolution is at least {\it locally} (in `time') meaningful.

Advancing to more generic situations in quantum cosmology quickly
leads to technical challenges such as constructing a
positive--definite inner product on the space of solutions to the
quantum constraints, known as the {\it Hilbert space problem}
\cite{pot,isham1}. In order to sidestep the {\it Hilbert space
problem} and extract qualitative and generic features from systems
otherwise too intricate to be solved exactly, effective techniques
have been developed \cite{martinbooks,bojski,effcon1,effcon2}. Based
on these techniques, an {\it effective approach to the problem of
time} has recently been introduced \cite{bht1,bht2} (see also
\cite{proc} for a brief summary) which allows us to evaluate the
relational quantum dynamics of systems featuring the {\it global
time problem} in the semiclassical regime. 
The effective approach as it stands applies to canonical algebras of basic quantum variables, and thereby, when applied to cosmology, it provides an effective description of Wheeler--DeWitt (WDW) quantum cosmology.
This effective approach
seems especially well suited for analyzing semiclassical dynamics
of nonintegrable systems, because two of its main achievements are
firstly, to make sense of (temporally) local time evolution with
(temporally) local relational observables, and secondly, to cope
with imperfect clocks by allowing one to explicitly switch back and
forth between different internal times, thereby avoiding clock
pathologies. In this manner---and in analogy to local coordinates on
a manifold---one is enabled to cover semiclassical evolution
trajectories by patches of local relational times. In \cite{bht2},
this effective approach was applied to two simple toy models with
decoupled clocks.

It is the goal of the present article to take a step away from
deparametrizations with `ideal clocks,' making a first step towards
the generic situation by considering (more realistically) coupled
clock degrees of freedom in a nonintegrable cosmological model.
Concretely, although observationally a flat universe seems to be
favored \cite{wmap}, we shall investigate the closed Friedman--Robertson--Walker (FRW) model
filled with a minimally coupled massive scalar field in order to
specifically address the issue of relational evolution. This model
universe has been studied extensively in the literature
\cite{hawk1,hawk2,page,page2,Kiefer,belinsky,kkt,starobinsky,cornshell,lafshell,zehbook,singh},
in particular because it constitutes a simple cosmology which
`generically' produces inflation.
While the classical dynamics of this system are understood in detail \cite{page,belinsky,kkt,cornshell}, its complete and consistent quantization is still pending in any approach to quantum cosmology. The troubles in constructing a complete quantization are rooted in the classical nonintegrability and the absence of a (temporally) global internal clock, which leads to nonunitarity and thus far impeded discussion of relational evolution. In the main body of this article, we shall explain some of the quantum troubles and---at least in the semiclassical regime---make some headway as regards relational evolution in this model universe by means of the {\it effective approach to the problem of time}. 
Whereas the resolution of the classically singular region through a quantum bounce in effective loop quantum cosmology (LQC) was studied in \cite{singh},\footnote{Singularity avoidance in this model within the framework of semiclassical gravity was earlier reported in \cite{parker}.} in the effective WDW formulation investigated here, we will rather focus on the region of maximal expansion which features a chaotic scattering and is thus especially challenging for relational dynamics. Although this work relies on approximating the WDW quantization of
the model, our discussion may also be relevant for the construction
of relational observables in LQC, especially since the chaotic
behavior we study occurs for large values of the scale factor, where
LQC approximately reproduces the WDW formulation \cite{lqc}.

Attention will be devoted to conceptual issues raised in the
earlier literature as regards the initial value problem and the
semiclassical limit \cite{zeh,Kiefer,zehbook}. The primary result of
the present work is strong evidence that quantum relational
evolution in this model, while possible for sufficiently
semiclassical states, generically breaks down in the region of
maximal expansion; nonintegrability leads to a defocusing of
nearby classical trajectories and thereby to a breakdown of
semiclassicality.
In addition, the chaotic behavior of the model can lead to a complicated
structure of phase space orbits on all scales, making it
fundamentally impossible to construct semiclassical states peaked
around a large class of classical orbits.
These results shed a first light on (the breakdown of) relational
quantum evolution in generic situations and highlight the nontrivial
nature of the question posed in the beginning of this section.

The rest of this manuscript is organized as follows: For the
convenience of the reader, Sec.~\ref{sec_eff} reviews elements of
effective techniques for constrained quantum systems. Next, in
Sec.~\ref{sec_semeff}, the effective semiclassical truncation of a
general class of two--component dynamical systems, applicable to
homogeneous cosmology, is detailed. In particular, the general
construction for switching clocks in such systems is provided.
Subsequently, in Sec.~\ref{sec_frw}, we examine in detail the closed
FRW model universe minimally coupled to a massive scalar field and
finally conclude with a discussion and an outlook in
Sec.~\ref{sec_con}.

\section{Effective equations for constrained quantum systems}\label{sec_eff}

The idea behind the effective approach is to avoid operating with
specific Hilbert space representations and, instead, to focus on
extracting representation--independent information.
Here we focus on two--component quantum systems with a single
constraint that, in addition, plays the role of the Hamiltonian, as
is appropriate for homogeneous cosmology with a scalar field providing
the matter content. For the purposes of this section, we label the
two components by their respective coordinate and momentum operators
satisfying the canonical commutation relations:
\[
[ \hat{q}_1, \hat{p}_1 ] = i\hbar , \quad [ \hat{q}_2, \hat{p}_2 ] =
i\hbar \,.
\]
In the effective approach, we describe a quantum state through the values it assigns to the
four \emph{expectation values} $\langle \hat{q}_1 \rangle$, $\langle
\hat{p}_1 \rangle$, $\langle \hat{q}_2 \rangle$, $\langle \hat{p}_2
\rangle$\ and the (countably) infinite set of \emph{moments} \cite{bojski,effcon1,effcon2,bht2},
\begin{eqnarray}\label{eq:def_moments}
 \Delta(q_1^{a}p_1^{b} q_2^{c}p_2^{d})&:=& \langle(\hat{q}_1-\langle\hat{q}_1\rangle)^{a}
(\hat{p_1}-\langle\hat{p_1}\rangle)^{b} 
(\hat{q}_2-\langle\hat{q}_2\rangle)^{c}
(\hat{p_2}-\langle\hat{p_2}\rangle)^{d} \rangle_{\rm Weyl}\,, \nn
\end{eqnarray}
defined for $(a+b+c+d)\geq2$, where the latter quantity will be
referred to as the \emph{order} of a given moment. The subscript
``Weyl'' indicates totally symmetrized ordering of the product of
operators inside the brackets. The space coordinatized by the
expectation values and moments carries a natural phase space
structure defined by the Poisson bracket \ba\label{poisson}
 \{\langle\hat{A}\rangle,\langle\hat{B}\rangle\}=
\frac{\langle[\hat{A},\hat{B}]\rangle}{i\hbar}
\ea
for any pair of operators $\hat{A}$ and $\hat{B}$, extended to the
moments using the Leibniz rule and linearity.  If there is a true
Hamiltonian, it quickly follows from the Heisenberg equation that
the evolution of expectation values and moments is generated by the
Hamiltonian flow of the quantum Hamiltonian function
$H_Q(\langle\hat{q}_1\rangle, \langle\hat{p}_1\rangle,
\langle\hat{q}_2\rangle, \langle\hat{p}_2\rangle; \Delta(\cdots))=
\langle\hat{H}\rangle$ \cite{bojski,effcon1}.

For a system with a single constraint represented by an operator
$\hat{C}$, we follow Dirac's constraint quantization condition
and demand that physical states satisfy $\hat{C} |\psi \rangle =0$.
The analogue of this condition has been formulated directly on the
expectation values in \cite{effcon1,effcon2} as
\begin{equation} \label{eq:exvalconstraint}
\langle \widehat{pol} \hat{C} \rangle = 0
\end{equation}
for all polynomials $\widehat{pol}$\ in the four basic variables.
Intuitively, this corresponds to eliminating all of the quantum
modes involving the constraint operator. The constraint conditions
can be systematically imposed by using a linear basis for the
polynomial algebra. This leads to an infinite set of constraint
conditions on the space of expectation values and moments, which are
moreover first class with respect to the quantum Poisson bracket
defined earlier; i.e., the bracket between any two constraint
functions vanishes when~(\ref{eq:exvalconstraint}) is satisfied \cite{effcon1,effcon2}.
This, in particular, means that the above constraints induce
\emph{quantum gauge transformations} on the space of solutions
to~(\ref{eq:exvalconstraint}) via their Hamiltonian flows. It is
easy to see directly from the definition of the quantum Poisson
bracket that these flows only affect the expectation values of
operators whose quantum commutators with the constraint have a
nonvanishing expectation value on the quantum constraint surface;
such operators do not correspond to the true physical degrees of
freedom of the system (also known as the \emph{Dirac observables}).
In this formulation, the search for the physical observables is then
replaced by the search for functions of expectation values of
polynomial operators, which are invariant along the quantum Poisson
flows generated by the constraint
functions of~(\ref{eq:exvalconstraint}) along the constraint surface
they define.

Reducing the kinematical system
by the action of the constraint is
not practically feasible at this step:
we are dealing with an infinite dimensional quantum phase space,
where we need to impose an infinite set of constraint conditions
given by~(\ref{eq:exvalconstraint}) and integrate all of the
corresponding gauge flows. Fortunately, the system may be
approximately represented by a finite number of degrees of freedom
in the semiclassical regime with the help of the moments defined
above in~(\ref{eq:def_moments}). Specifying the values of the four
expectation values and all of the moments is entirely equivalent to
specifying the expectation values of all symmetrized products of the
four basic variables: we elect to use the moments, as they follow a
clear hierarchy when evaluated in a semiclassical state. In
particular, we assume that a moment of order $N$\ is of the same
semiclassical order as $\hbar^{N/2}$\ and approximate the system by
truncating both the degrees of freedom and the system of constraints
at some finite order in the semiclassical expansion. This hierarchy
is explicitly realized for a class of Gaussian wave functions in an
ordinary Schr\"odinger representation of a quantum
particle~\cite{Brizuela} but also holds in a more general class of states.
Further details of the effective framework will be explained along the way.

\section{Leading--order quantum corrections and effective dynamics}\label{sec_semeff}

In the present work we restrict our attention to classical
Hamiltonian constraints of the form \ba\label{ccon} C_{class}=p_1^2
- p_2^2 - V(q_1, q_2) \,, \ea where $V(q_1, q_2)$\ is polynomial, or
at least has a convergent power series expansion in $q_1$\ and
$q_2$. This class of Hamiltonian constraints covers several
homogeneous cosmological models, one of which is studied in detail in
Sec.~\ref{sec_frw}. Since no terms involve products of
noncommuting variables, we take the corresponding constraint
operator to be \ba\label{qcon} \hat{C} = \hat{p}_1^2 - \hat{p}_2^2 -
V(\hat{q}_1, \hat{q}_2) \,. \ea We systematically impose the
constraint conditions of~(\ref{eq:exvalconstraint}) by demanding
\ba\label{cond} \langle \hat{q}_1^a\hat{p}_1^b\hat{q}_2^c\hat{p}_2^d
\hat{C} \rangle = 0\, \ea for all non-negative integer values of
$a$, $b$, $c$, $d$. We focus on the leading--order quantum
corrections, which corresponds to truncating the system above
semiclassical order $\hbar$. Up to this order the kinematics of our
system is described by 14 independent functions: four
expectation values of the form $a= \langle \hat{a} \rangle \propto
\hbar^0$, four spreads of the form $(\Delta a)^2 = \langle
(\hat{a}-a)^2 \rangle\propto \hbar$, and six covariances of the form
$\Delta(ab)= \langle(\hat{a}-a)(\hat{b}-b)\rangle_{\rm Weyl} \propto
\hbar$. Note that, due to symmetrization, $\Delta(ab) = \Delta(ba)$,
and $a$\ is used to label both the classical function and the
expectation value of the corresponding quantum operator
$\hat{a}$---it should be clear from the context which of the above
it represents. We will use this notation throughout the rest of the
present work. After the truncation, five nontrivial independent
constraint functions remain
(obtained via the substitution $\hat{a}=a+(\hat{a}-a)$ and
Taylor--expanding (\ref{cond}) about the expectation values
\cite{bojski,effcon1,effcon2,bht2}):
\begin{eqnarray}\label{eq:deparam_effC} C :=\q\,\,\,\,
\langle \hat{C}\rangle\q\q&=&p_1^2 - p_2^2+ (\Delta p_1)^2 - (\Delta
p_2)^2- V - {\textstyle\frac{1}{2}} \ddot{V}(\Delta q_1)^2  -
{\textstyle\frac{1}{2}} V''(\Delta q_2)^2 - \dot{V}'\Delta(q_1q_2),
\nonumber
\\ C_{q_1} := \langle (\hat{q}_1 - q_1) \hat{C}\rangle &=& 2p_1\Delta(q_1p_1)+ i\hbar
p_1 - 2p_2 \Delta(q_1p_2)-\dot{V} (\Delta q_1)^2 -
V' \Delta(q_1q_2),\nn\\
C_{p_1}:= \langle (\hat{p}_1 - p_1) \hat{C}\rangle &=& 2p_1(\Delta p_1)^2 -
2p_2\Delta(p_1p_2)-
\dot{V}(\Delta(q_1p_1)-{\textstyle\frac{1}{2}}i\hbar)
- V' \Delta(p_1q_2),\\
C_{q_2}:= \langle (\hat{q}_2 - q_2) \hat{C} \rangle &=& 2 p_1
\Delta(p_1q_2)- 2p_2 \Delta(q_2p_2)- i\hbar p_2 - \dot{V}
\Delta(q_1q_2) -
V'(\Delta q_2)^2,\nn\\
C_{p_2}:= \langle (\hat{p}_2 - p_2) \hat{C} \rangle &=& 2 p_1 \Delta
(p_1 p_2) - 2p_2 (\Delta p_2)^2 - \dot{V} \Delta(q_1p_2) -
V'(\Delta(q_2p_2)-{\textstyle\frac{1}{2}}i\hbar) \,.\nn
\end{eqnarray}
Here and
in the rest of the present section,
we will use the shorthand notation where dots over $V$\ denote
partial derivatives with respect to $q_1$, primes denote partial
derivatives with respect to $q_2$, and we drop explicit reference to
the arguments, so that e.g.\ $\dot{V} = \frac{\partial V}{\partial
q_1} (q_1, q_2)$.
The system of constraint functions is simple to solve; however the
Poisson flows they generate are, in general, difficult to integrate
and interpret. The flows have the following general feature: on the
constraint surface, a nontrivial combination of constraints
$C_{q_1}, C_{p_1}, C_{q_2}, C_{p_2}$\ has a vanishing flow. This
happens due to the degeneracy of the Poisson structure \cite{bojski,effcon1,effcon2,bht2} and is
important for the correct reduction in the degrees of freedom.
Following the method which was suggested in \cite{effcon1},
formalized in \cite{effcon2} and
conceptually more thoroughly elucidated in \cite{bht1,bht2},
we partially fix the gauge freedom by choosing one of the configuration variables as an internal clock and
interpret the single remaining quantum flow as the dynamics.
Choosing $q_1$\ as the clock, we impose three `$q_1$--gauge' conditions, in order to `project the relational clock $q_1$ to a classical parameter' \cite{bht1,bht2}:
\begin{eqnarray}\label{eq:q1gauge}
\phi_1 := (\Delta q_1)^2 = 0,\q\quad \phi_2:=\Delta(q_1q_2) =
0 ,\q\q \phi_3 := \Delta(q_1p_2) = 0. 
\end{eqnarray}
In fact, the gauge conditions essentially determine to which Hilbert space representation and clock time slicing (in a deparametrization) the effective relational evolution will correspond \cite{bht2}.\footnote{The quantum phase space, by construction, includes a very general
collection of linear functionals on the polynomial algebra of the
kinematical coordinate--momentum variables. It therefore contains the
information about a general class of (in general inequivalent) Hilbert
space representations of this algebra based on slicings in a (local)
deparametrization (see Sec.\ IV C in \cite{bht2} for a detailed discussion).}
One gauge flow remains, which preserves both the constraints and the
above gauge conditions and is
generated by the `Hamiltonian' constraint (see \cite{bht2} for details on how to obtain $C_H$):
\begin{equation}\label{eq:effCham_q1}
C_H:=C-\frac{1}{2p_1}C_{p_1} - \frac{p_2}{2p_1^2}C_{p_2} -
\frac{V'}{4p_1^2}C_{q_2}\,.
\end{equation}
We solve the constraint functions by eliminating $p_1$\ and the
moments generated by $\hat{p}_1$. The remaining degrees of freedom
are captured by the moments and expectation values of $\hat{q}_2$\
and $\hat{p}_2$\ only, i.e.\ $q_2, p_2, (\Delta q_2)^2,
\Delta(q_2p_2), (\Delta p_2)^2$, as well as the expectation value
$q_1$. We interpret the resulting system as expectation values and
moments generated by the pair $\hat{q}_2$, $\hat{p}_2$\ evolving
relative to the internal clock $q_1$,
where the corresponding equations of motion are obtained through the Poisson structure (\ref{poisson}).\footnote{Expectation values satisfy the classical Poisson brackets (as is obvious from (\ref{poisson})) and commute with the moments. Furthermore, the Poisson algebra of the moments of two canonical pairs can be found in the appendices of \cite{effcon2,bht2}. In this article, explicit use is made of this Poisson structure wherever equations of motion are calculated.}
For consistency of this
interpretation we require that the values of these variables satisfy
\emph{positivity} conditions
\begin{eqnarray}
&& q_2, p_2, (\Delta q_2)^2, (\Delta p_2)^2, \Delta(q_2p_2) \in \mathbb{R}, \nonumber \\
&& (\Delta p_2)^2, (\Delta q_2)^2 \geq  0, \nonumber \\ && (\Delta
q_2)^2 (\Delta p_2)^2 - \left(\Delta(q_2p_2)\right)^2 \geq
\frac{1}{4} \hbar^2\q. \label{eq:pos_conditions1}
\end{eqnarray}
In \cite{bht1,bht2} it was found that the expectation value of the clock picks up a specific imaginary contribution
\begin{equation}\label{imt}
\Im[q_1]= -\frac{\hbar}{2p_1}\,
\end{equation}
in order for the constraint $C_H$\ of~(\ref{eq:effCham_q1}) to be consistently satisfied and for the evolving variables to remain real along the flow generated
by it.
The real part of $q_1$\ can be used to parametrize the evolution
flow, which preserves the above form of the imaginary contribution
to $q_1$. The gauge-fixing conditions together with the
inequalities of~(\ref{eq:pos_conditions1})
and the interpretation of the remaining flow as the evolution in (the
real part of) the internal clock $q_1$\ will from now on be referred
to as the \emph{Zeitgeist} associated with $q_1$.
(Temporally local) relational observables---or {\it fashionables}---are then given by the correlations of moments and expectation values with the expectation value of the clock and can be interpreted as describing an approximate (local) unitary evolution in $q_1$ \cite{bht1,bht2}. Notice that these fashionables are state dependent.
We can follow an
entirely analogous sequence of steps choosing $q_2$ as the clock,
with a minor difference due to the minus sign in front of
$\hat{p}_2^2$, yielding a slightly different expression for the new
evolution flow generator:
\begin{equation}\label{eq:effCham_q2}
C_H:=C-\frac{1}{2p_2}C_{p_2} - \frac{p_1}{2p_2^2}C_{p_1} +
\frac{\dot{V}}{4p_2^2}C_{q_1}\,.
\end{equation}
The gauge--fixing conditions, imaginary contribution to the clock and
positivity conditions are all obtained by simply switching the
labels `$1$'\ and `$2$.'

Notice that relational evolution in a chosen clock is not only most conveniently interpreted in the corresponding {\it Zeitgeist}, but, furthermore, in every {\it Zeitgeist} we evolve a {\it different} set of relational observables (see \cite{bht1} and especially Sec.\ IV C in \cite{bht2}).

\subsection{Failure of a {\it Zeitgeist} and transient observables}\label{sec_failzeit}

For the class of Hamiltonian constraints considered here, $q_1$\ is
in general not a globally valid clock along the gauge orbits. The
breakdown occurs when the evolution rate of the clock becomes very small or vanishes and
the clock ``reverses direction.'' Classically, this happens when
$\{q_1, C_{class}\} = 2p_1 = 0$, which is possible as $p_1$\ is in
general not a constant of motion in these systems. On the effective
side, as the expectation value $p_1$\ approaches zero, the $q_1$--\emph{Zeitgeist} together with its
physical interpretation becomes incompatible with the semiclassical
approximation. One can infer this already from the form of the
imaginary contribution to the clock~(\ref{imt}), which becomes
divergently large as $p_1$\ approaches zero. Furthermore, the equations of motion for the evolving moments
become singular at $p_1=0$, and the moments diverge as they are
evolved towards this singularity because the coefficients in the evolution
generator~(\ref{eq:effCham_q1}) also diverge as we approach a turning
point.

Intuitively, the clock will simply be too slow to resolve the
evolution of other degrees of freedom with respect to it when its
momentum becomes small (compared to the relevant scale in the
system) and will thereby lead to large fluctuations in the (relative to
$q_1$, fast) evolving degrees of freedom. These
fluctuations/uncertainties must diverge as the clock `stops' (and
thus becomes maximally `imperfect'). The important consequence is
that the quantum evolution in $q_1$ breaks down {\it before} the
classical turning point, and therefore the relational observables in
the $q_1$--\emph{Zeitgeist} are only locally valid \cite{bht1,bht2};
they are {\it transient observables}. Notice that the range of
validity of the \emph{Zeitgeist} crucially depends on the state.

This is the effective analogue of nonunitarity in $q_1$ evolution.
Indeed, as argued in \cite{bht2} by analogy with a Schr\"odinger
regime in $q_1$ internal time, a condition such as $(\Delta
q_1)^2=0$ (as required by the $q_1$--\emph{Zeitgeist}) is
inconsistent in the turning region:
violation of unitary evolution would generally result in loss of
normalization, so that $\langle \mathds{1}\rangle=1$\ will not be preserved
leading to a nonzero value for $(\Delta q_1)^2=\langle
q_1^2\rangle-\langle q_1\rangle^2=q_1^2\left(\langle
\mathds{1}\rangle-\langle  \mathds{1}\rangle^2\right)$.
The clock degree of freedom thus cannot be `projected to a
classical parameter' anymore and the interference of segments of the
wave function before and after the classical turning point causes a
mixing of internal time directions, i.e.\ of positive and negative
values of the clock momentum \cite{bht1,bht2}. This
conclusion
is in agreement with the analysis in
\cite{marolf1,marolf2,gidmarhar}, where it was shown that a good
resolution of relational observables and evolution requires the
clock to be essentially decoupled from the other degrees of freedom
and its momentum to be large.\footnote{On the other hand, large
energies (or momenta) are an intricate issue in gravitational
physics due to black hole formation. Consequently, there is a cap on the clock's energy and thus on the accuracy of physical clocks
\cite{gampul}.}
 In this situation, one recovers `good unitary quantum mechanics' in both the Schr\"odinger and Heisenberg picture from the relational dynamics \cite{marolf1,marolf2}. The state clearly plays a key role in the recovery of an `accurately--resolved unitary' evolution and, in fact, may entirely prevent it if it is highly quantum in nature \cite{marolf1,gidmarhar,bht1,bht2}.

Does this indicate that the state is no longer semiclassical past
such a turning point? Not necessarily---the semiclassical
assumption breaks only relative to a specific set of gauge
conditions, and the other configuration variable $q_2$\ can serve as
a good internal clock near a turning point of $q_1$, so long as
$\{q_2, C_{class} \} = -2p_2\neq 0$. We can select $q_2$\ as a clock
and follow through with the associated gauge fixing and construction
of relational observables in a manner entirely analogous to the one
employed when choosing $q_1$\ as time. This new $q_2$--gauge can
eventually also fail, at which point it may be safe to use $q_1$\ as
a clock again. It is precisely in order to emphasize this transient
nature of the above internal clock frameworks, that we refer to the
$q_1$\ gauge and its corresponding dynamical interpretation as the
$q_1$--\emph{Zeitgeist}. However, if we do not wish to commit to a
single clock, we need a method for transferring relational data
between the two gauge frameworks.

\subsection{Transformation to a different clock}\label{sec:general_transf}

In order to clarify what would constitute the desired gauge
transformation, we begin with a few remarks on the geometry of the
situation at hand. The two--component system is described at order
$\hbar$\ by 14 kinematical degrees of freedom. The truncated
system of constraints gives five functionally independent conditions
$C_i=0$\ on this space, which therefore restrict the system to a
nine--dimensional surface. Five constraint functions, in general,
generate four independent flows or vector fields $X_{C_i}$\ on this
surface through the Poisson bracket $X_{C_i} (f) = \{ f, C_i\}$,
which integrate to a four--dimensional gauge orbit. We have
introduced three partial gauge--fixing conditions $\phi_i=0$, e.g.\ (\ref{eq:q1gauge}), that
break three of the four gauge flows, such that only one
independent combination of the vector fields $X_{C_i}$\
preserves the gauge; we interpret this flow as the dynamics in the
relevant clock variable. Geometrically, these one--dimensional orbits
are formed by the intersection of the surface defined by the gauge
conditions $\phi_i = 0$\ with the integral orbits of the set of
vector fields $X_{C_i}$\ on the constraint surface. Surfaces
corresponding to a different set of gauge conditions $\phi'_i=0$\
associated with a different internal clock give different
one--dimensional intersections with the gauge orbits and, therefore,
a different evolution flow. In order to interchange the relational
data consistently, we need to go from the $\phi_i = 0$\ surface to the
one defined by $\phi'_i=0$\ without moving off of a given gauge
orbit. The most natural way to achieve this is to follow the gauge
flows themselves, i.e.\ to find a combination of the vector fields
$X_{C_i}$\ whose integral curve intersects both $\phi_i = 0$\ and
$\phi'_i=0$.

Let us be concrete now. Recall that the $q_1$--\emph{Zeitgeist} is given by
the conditions $(\Delta q_1)^2 = \Delta(q_1q_2) = \Delta(q_1p_2) =
0$. The last condition is equivalent to $\Delta(q_1p_1) =
-i\hbar/2$\ if we impose the constraints and the other two gauge
conditions, which can be seen directly from $C_{q_1}$\
in~(\ref{eq:deparam_effC}). In this section we will use this
alternative form of the third gauge condition for convenience.
Similarly, the $q_2$--gauge is given by $(\Delta q_2)^2 = \Delta(q_1q_2)
= \Delta(p_1 q_2) = 0$, where the last condition is equivalent to
$\Delta(q_2p_2) = -i\hbar/2$. To transform from $q_1$--gauge to
$q_2$--gauge we need to find a combination of the vector fields $G =
\sum_i \xi_i X_{C_i}$, such that a (possibly finite) integral of its
flow transforms the variables as
\begin{equation}
\left\{ \begin{array}{l} (\Delta q_2)^2 = (\Delta q_2)_0^2 \\
\Delta(q_1q_2) = 0 \\ \Delta(q_2p_2) = \Delta(q_2p_2)_0 \end{array}
\right. \rightarrow \left\{ \begin{array}{l} (\Delta
q_2)^2 = 0 \\ \Delta(q_1q_2) = 0 \\
\Delta(q_2p_2) = -i\hbar/2 \end{array} \right. \ ,
\label{eq:tgauge_to_qgauge}
\end{equation}
where the subscript `$0$' labels the value of the corresponding
variable prior to the gauge transformation.

In general, one would expect such a transformation to be unique up
to the dynamical flows of the two
`Hamiltonian' constraints in the respective \emph{Zeitgeister},
since they preserve the
corresponding sets of gauge conditions. To fix this freedom, and to
make the transformation induced on the expectation values small, we
fix the multiplicative coefficient of $X_C$\ in $G$\ to zero. There
is still some freedom in choosing a path for the gauge
transformation: the five constraints generate only four independent
flows. Removing $C$ still leaves us with three independent flows
which we can combine. At this point we construct the gauge
transformation in two steps. First we search for a flow that
satisfies $G_1\left(\Delta(q_2p_2) \right) = G_1\left(
\Delta(q_1q_2) \right)=0$\ on the constraint surface and rescale
the flow such that $G_1\left( (\Delta q_2)^2 \right) = 1$. The
second step involves finding the flow that satisfies
$G_2\left((\Delta q_2)^2 \right) = G_2\left( \Delta(q_1q_2)
\right)=0$ and rescaling this flow such that $G_2\left(
\Delta(q_2p_2) \right) = 1$. The required gauge transformation will
then be given by integrating the flow along $G = -(\Delta q_2)^2_0
G_1-(\Delta(q_2p_2)_0+i\hbar/2)G_2$.

The condition $\Delta(q_1q_2) = 0$\ is shared by both gauge choices
and is preserved by $G$\ by construction. We will therefore use this
condition to simplify the form of the gauge--transformation fields
$G_1$\ and $G_2$. The conditions we have imposed determine $G_1$\
and $G_2$\ uniquely, and after a number of algebraic manipulations,
some of which were performed with the aid of MATHEMATICA 7, one
obtains the explicit effect of $G_1$\ and $G_2$\ on the free
variables of the $q_2$--gauge:
\begin{eqnarray*}
G_1(q_1)= -\frac{p_1\dot{V}+2p_2V'}{4p_1p_2^2} \quad&,&\quad G_2(q_1)= -\frac{1}{p_1}\,,\\
G_1(p_1)= -\frac{p_1\ddot{V}+p_2\dot{V}'}{4p_2^2} \quad&,&\quad G_2(p_1)= 0\,,\\
G_1(q_2) = \frac{V'}{4p_2^2} \quad&,&\quad G_2(q_2) = \frac{1}{p_2}\,,\\
G_1(p_2)= -\frac{p_1\dot{V}'+p_2V''}{4p_2^2} \quad&,&\quad
G_2(p_2)=0 \,,
\end{eqnarray*}
\begin{eqnarray*}
G_1 \left( (\Delta q_1)^2 \right) =
-\frac{p_1^2}{p_2^2} \quad&,&\quad G_2 \left( (\Delta q_1)^2 \right)
= 0 \,,\\ G_1 \left( (\Delta p_1)^2 \right) = -\dot{V}
\frac{p_1\dot{V}+2p_2V'}{4p_1p_2^2} \quad&,&\quad G_2 \left( (\Delta
p_1)^2 \right) = -\frac{\dot{V}}{p_1}\,,\\
G_1 \left( \Delta(q_1p_1) \right) = -\frac{p_1\dot{V} + p_2V'}{2p^2}
\quad&,& G_2 \left( \Delta(q_1p_1) \right) = -1\,.
\end{eqnarray*}
By inspecting~(\ref{eq:tgauge_to_qgauge}), we infer that, in order to
transform between the two gauges, we need to follow the integral
curve of the vector field $G$\ for an interval of the flow parameter
equal to 1. We denote the flow of $G$\ by $\alpha_G^s$, with flow parameter $s$. Scalar functions transform via
dragging their argument along the flow as
$\alpha_G^s.f(x)=f\left( \alpha_G^s(x) \right)$, $x\in\cc$ where $\cc$ is the constraint surface. The family of
translated functions varies differentiably along the flow according
to the equation
\begin{equation}
\frac{d}{ds} \left( \alpha_G^s.f \right) (x)= G \left( \alpha_G^s.f
\right) (x)\ .
\end{equation}
If $f(x)$\ is smooth along $G$, the solution to the above equation
can be constructed through the derivative power series
\begin{equation}
\alpha_G^s.f (x) = \sum_{n=0}^{\infty} \frac{s^n}{n!} G^n(f)(x) \ ,
\end{equation}
where $G^n(f)$\ is the $n$th derivative of $f$\ along $G$, i.e.\
$G^n(f) = G\left( G^{n-1} (f) \right)$\ with $G(f)$\ defined as
usual.

Here, we are only interested in the transformations to order
$\hbar$. In our case $s=1$\ and $G = aG_1+bG_2$, where $a$\ and $b$\
are constants of order $\hbar$. In addition, for all expectation
values and moments, $G_1(f)$\ and $G_2(f)$\ are of classical order.
It follows that in the series solution for finite gauge
transformations the terms proportional to the second derivative
along $G$\ and higher will be of order above $\hbar$. We can
therefore approximate the gauge transformation to the desired order
by the leading--order terms, i.e.\
\begin{equation}
\alpha_G^1.f (x) = f(x) + G(f)(x) + o(\hbar^2)\ .
\end{equation}

The evolving variables in the $q_2$--\emph{Zeitgeist} (appearing on the
left-hand-side and labeled by the subscript ``new'') in terms of
those in the $q_1$--gauge (appearing on the right-hand-side) are given
by
\begin{eqnarray}\label{ch5:eq:q1toq2}
q_{1\ {\rm new}} &=& q_1 + \frac{i\hbar}{2p_1} +
\frac{p_1\dot{V}+2p_2V'}{4p_1p_2^2} (\Delta q_2)^2 + \frac{1}{p_1}
\Delta(q_2p_2)\nn
\\ q_{2\ {\rm new}} &=& q_2 -\frac{i\hbar}{2p_2} - \frac{V'}{4p_2^2}
(\Delta q_2)^2 - \frac{1}{p_2} \Delta(q_2p_2)\nn \\
p_{1\ {\rm new}} &=& p_1 + \frac{p_1 \ddot{V} + p_2\dot{V}'}{4p_2^2}
(\Delta q_2)^2\nn \\ p_{2\ {\rm new}} &=& p_2 +
\frac{p_1\dot{V}'+p_2V''}{4p_2^2}(\Delta q_2)^2\\
(\Delta q_1)^2_{\rm new} &=& \frac{p_1^2}{p_2^2} (\Delta q_2)^2\nn \\
\Delta(q_1p_1)_{\rm new} &=& \Delta(q_2p_2) + \frac{p_1\dot{V} + p_2
V'}{2p_2^2} (\Delta q_2)^2\nn \\ (\Delta p_1)^2_{\rm new} &=&
\frac{1}{p_1^2 p_2^2} \left[ p_2^4 (\Delta p_2)^2 + p_2^2
(p_1\dot{V}+p_2V')\Delta(q_2p_2) + \frac{1}{4} (p_1\dot{V}+p_2V')^2
(\Delta q_2)^2\right]\nn
\end{eqnarray}
All other variables are either gauge fixed or eliminated using the
second--order constraint functions
$C_{q_1}$, $C_{q_2}$, $C_{p_1}$, $C_{p_2}$.
The reverse transformation---obtained in an entirely analogous
manner---is given by:
\begin{eqnarray}
q_{1\ {\rm new}} &=& q_1 -\frac{i\hbar}{2p_1} +
\frac{\dot{V}}{4p_1^2} (\Delta q_1)^2 - \frac{1}{p_1} \Delta(q_1p_1)
\nn\\ q_{2\ {\rm new}} &=& q_2 +\frac{i\hbar}{2p_2} - \frac{2p_1\dot{V}
+ p_2V'}{4p_1^2p_2} (\Delta q_1)^2 + \frac{1}{p_2} \Delta(q_1p_1)\nn \\
p_{1\ {\rm new}} &=& p_1 - \frac{p_1 \ddot{V} + p_2\dot{V}'}{4p_1^2}
(\Delta q_1)^2\nn \\ p_{2\ {\rm new}} &=& p_2 -
\frac{p_1\dot{V}'+p_2V''}{4p_1^2}(\Delta q_1)^2\\
(\Delta q_2)^2_{\rm new} &=& \frac{p_2^2}{p_1^2} (\Delta q_1)^2\nn \\
\Delta(q_2p_2)_{\rm new} &=& \Delta(q_1p_1) - \frac{p_1\dot{V} + p_2
V'}{2p_1^2} (\Delta q_1)^2\nn \\ (\Delta p_2)^2_{\rm new} &=&
\frac{1}{p_1^2 p_2^2} \left[ p_1^4 (\Delta p_1)^2 - p_1^2
(p_1\dot{V}+p_2V')\Delta(q_1p_1) + \frac{1}{4} (p_1\dot{V}+p_2V')^2
(\Delta q_1)^2\right]\nn
\end{eqnarray}
As expected, the two transformations invert each other up to terms
of order $\hbar^{3/2}$. It is also straightforward to verify that
these gauge transformations preserve, or rather transform, the
positivity conditions~(\ref{eq:pos_conditions1}). Consider, for
example, the transformation from $q_1$--\emph{Zeitgeist} to
$q_2$--\emph{Zeitgeist}, given by~(\ref{ch5:eq:q1toq2}). If the
values of the relational observables of the $q_1$--gauge satisfy
positivity, we can derive the following inequality (see Appendix B
of \cite{bht2}) for all $\alpha, \ \beta \in \mathbb{R}$:
\begin{equation}\label{eq:inequality}
\alpha^2 (\Delta q_2)^2 + \beta^2 (\Delta p_2)^2 + 2\alpha \beta
\Delta(q_2p_2) \geq 0.
\end{equation}
We quickly infer the following for the relational observables of the
$q_2$--gauge:
\begin{itemize}

\item The evolving variables (and thus the relational observables) in the $q_2$--\emph{Zeitgeist} are real, while $q_2$\ acquires an
imaginary contribution consistent with~(\ref{imt}).
\item $(\Delta q_1)^2\geq 0$\ follows immediately and $(\Delta
p_1)^2\geq 0$\ follows directly from the
inequality of~(\ref{eq:inequality}).

\item The generalized uncertainty relation $(\Delta q_1)^2 (\Delta p_1)^2
- \left(\Delta(q_1p_1)\right)^2 \geq \frac{\hbar^2}{4}$\ also
follows from~(\ref{eq:inequality})
after some algebraic manipulations (for more details, see the
Appendix of~\cite{bht2}).
\end{itemize}
Importantly, we see that the gauge transformations are entirely
consistent with the imaginary contribution to the expectation value
of the clock discussed earlier in this section.

The above construction provides us with the general equations to translate between different \emph{Zeitgeister} in the class of models described by (\ref{qcon}). Notice that since relational evolution in a given clock choice is best described in its corresponding \emph{Zeitgeist}, a change of clock thus necessitates a change of gauge \cite{bht2}. As shown in \cite{bht2}, the precise moment of switching between \emph{Zeitgeister} is irrelevant, so long as the two \emph{Zeitgeister} are valid before and after the gauge transformation, respectively.
The latter condition is crucial and will generally fail in the region of maximal expansion in the cosmological model to be studied in the sequel and thereby cause a breakdown of relational evolution.

\section{The closed FRW model universe minimally coupled to a massive scalar field}\label{sec_frw}

We now wish to extend the scope of the effective framework,
described in Secs.~\ref{sec_eff} and~\ref{sec_semeff}, by applying
it to quantum cosmology. We begin by discussing the relevant
classical features of the closed FRW model filled with a massive
scalar field in Sec.~\ref{sec_cl}, proceed by explaining troubles in
the quantization of this model in Sec.~\ref{sec_qt}, and examine the
effective dynamics in detail in Sec.~\ref{sec_frweff}.

\subsection{The classical dynamics}\label{sec_cl}

The action of a homogenous massive scalar field $\phi(t)$ minimally
coupled to a (homogeneous and isotropic) closed
Friedman--Robertson--Walker spacetime, of topology
$\mathbb{R}\times\mathbb{S}^3$ and described by the metric
\ba\label{metric} ds^2=-N^2(t)\,dt^2+a^2(t)\,d\Omega^2 \ea (where
$d\Omega^2$ is the line element on a unit $\mathbb{S}^3$), is given
by \ba\label{action} S[a,\phi]=&\frac{1}{2}\int dt\,Na^3
\left(-\left(\frac{1}{aN}\frac{da}{dt}\right)^2+\frac{1}{a^2}
+\left(\frac{1}{N}\frac{d\phi}{dt}\right)^2-m^2\phi^2\right)\q. \ea
Variation of the action with respect to lapse $N$, field $\phi$, and
scale factor $a$ yields the Friedman, `Klein--Gordon,' and
Raychaudhuri equation, respectively
(here we use notation
$\dot{}=N^{-1}\frac{d}{dt}$):
 \ba\label{lag}
&&\dot{a}^2=-1+a^2\left(\dot{\phi}^2+m^2\phi^2\right)\q,\label{fried}\\
&&\ddot{\phi}+\frac{3\dot{a}}{a}\dot{\phi}+m^2\phi=0\q,\label{KG}\\
&&\ddot{a}=a\left(m^2\phi^2-2\dot{\phi}^2\right)\q.\label{ray} \ea
These equations of motion are clearly not all independent (e.g.,
differentiating (\ref{fried}) and combining it with (\ref{KG}) gives
the Raychaudhuri equation (\ref{ray})).

Despite the apparent simplicity, the model possesses a surprisingly
rich solution space
\cite{hawk1,page,page2,belinsky,kkt,starobinsky,cornshell}. We do
not intend to review the details here, but wish to summarize and
pinpoint those classical aspects which are essential for our
subsequent discussion in the quantum theory.

This model universe attracted significant interest, mainly because
the mass term of the scalar field can act as an `effective
cosmological constant' in certain regimes and thereby drive a
deSitter--type inflationary period. Indeed, various phases of
cosmological evolution are possible because the equation of state of
the scalar field itself varies throughout evolution
\cite{belinsky,muk}. In \cite{belinsky} it was shown, using methods
of dynamical systems, that inflationary stages are a `generic'
property of solutions to (\ref{fried}, \ref{KG}). Setting initial
conditions at some small value of the scale factor, the scalar field
$\phi(t)$ decreases with increasing $a$, generating an inflationary
phase, and subsequently evolving to its equilibrium value
$\phi\approx0$, around which the field begins to
oscillate\footnote{As discussed in \cite{Kiefer,hawk2}, a solution
which expands out to a length scale on the order of $10^{60}$ Planck
lengths requires at least $10^{60}$ such oscillations of $\phi$.}
with frequency $m$, and the model universe exhibits a
matter--dominated era in which $a\propto t^{2/3}$
\cite{hawk1,page,page2,Kiefer,belinsky}. (The inflationary period is
longer for larger initial values $\phi_0$ of the scalar field
\cite{hawk1}.) Thereupon, the scale factor can begin to oscillate
between points of regular (nonglobal) maxima $a_{max,k}$ and
(nonglobal) minima $a_{min,k}$ \cite{page, page2,belinsky,kkt}. A
generic solution will evolve to a point of maximum extension---the
turning point---$a_{max}$, possibly oscillate around this point a
few times, and eventually recollapse to a big crunch singularity
\cite{hawk1,page,belinsky}. Thus, clearly, both $\phi$ and $a$ will
generically fail to be globally valid internal clock functions in
this model.\footnote{For small masses $m$, the scalar field
$\phi(t)$ is still a monotonically increasing function of $t$ as in
the massless case and thus a good global clock (see also the
discussion in \cite{kkt} in particular, the region in configuration
space called `region 0').}
Two typical classical solutions are displayed in
Fig.~\ref{fig:class_sol}.
\newcommand{\goodgap}{
\hspace{\subfigtopskip}
\hspace{\subfigbottomskip}}
\begin{figure*}[htp]
  \begin{center}
    \subfigure[]{\label{fig:class_sol-a}\includegraphics[scale=.55]{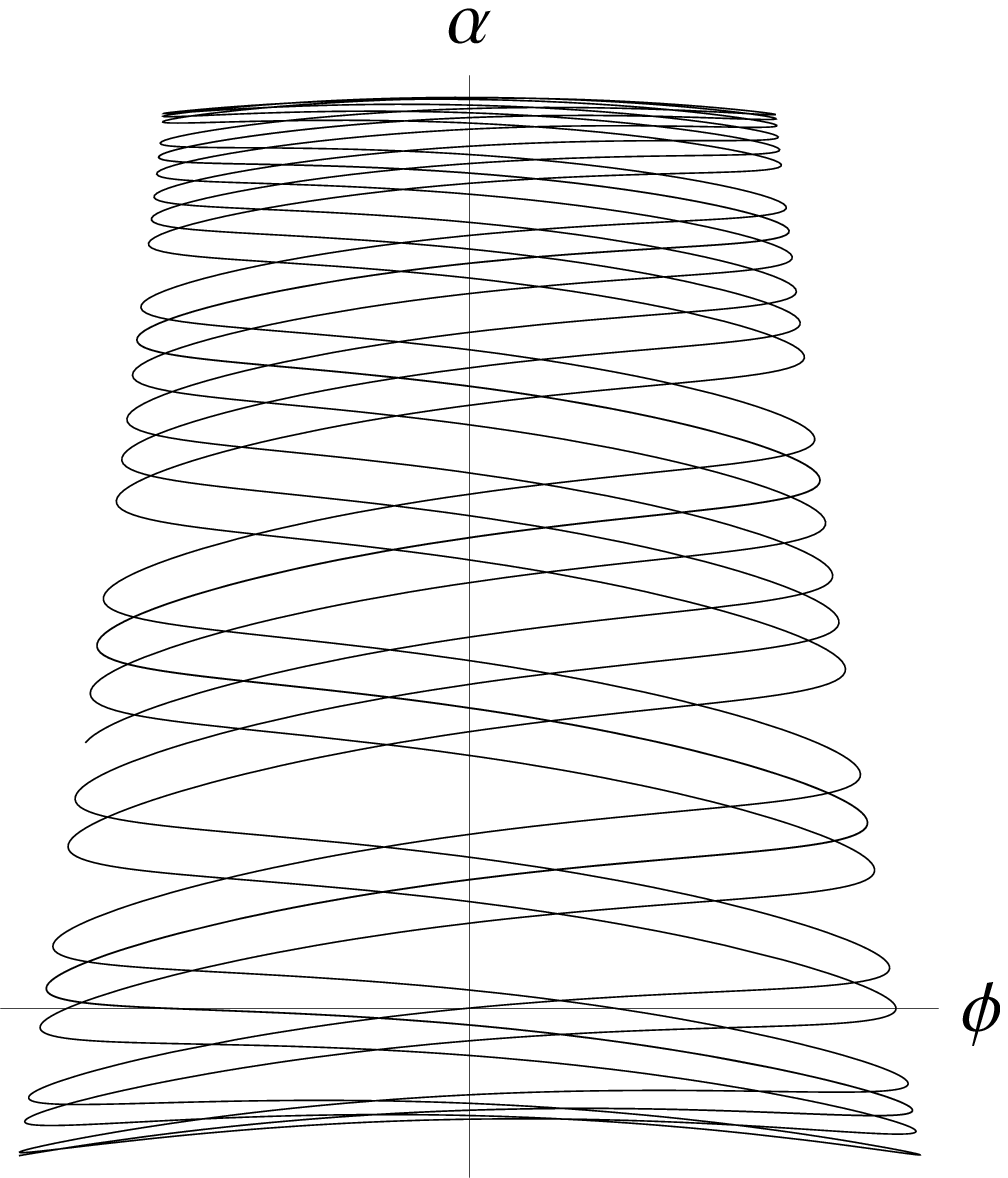}}\goodgap
    \subfigure[]{\label{fig:class_sol-b}\includegraphics[scale=.55]{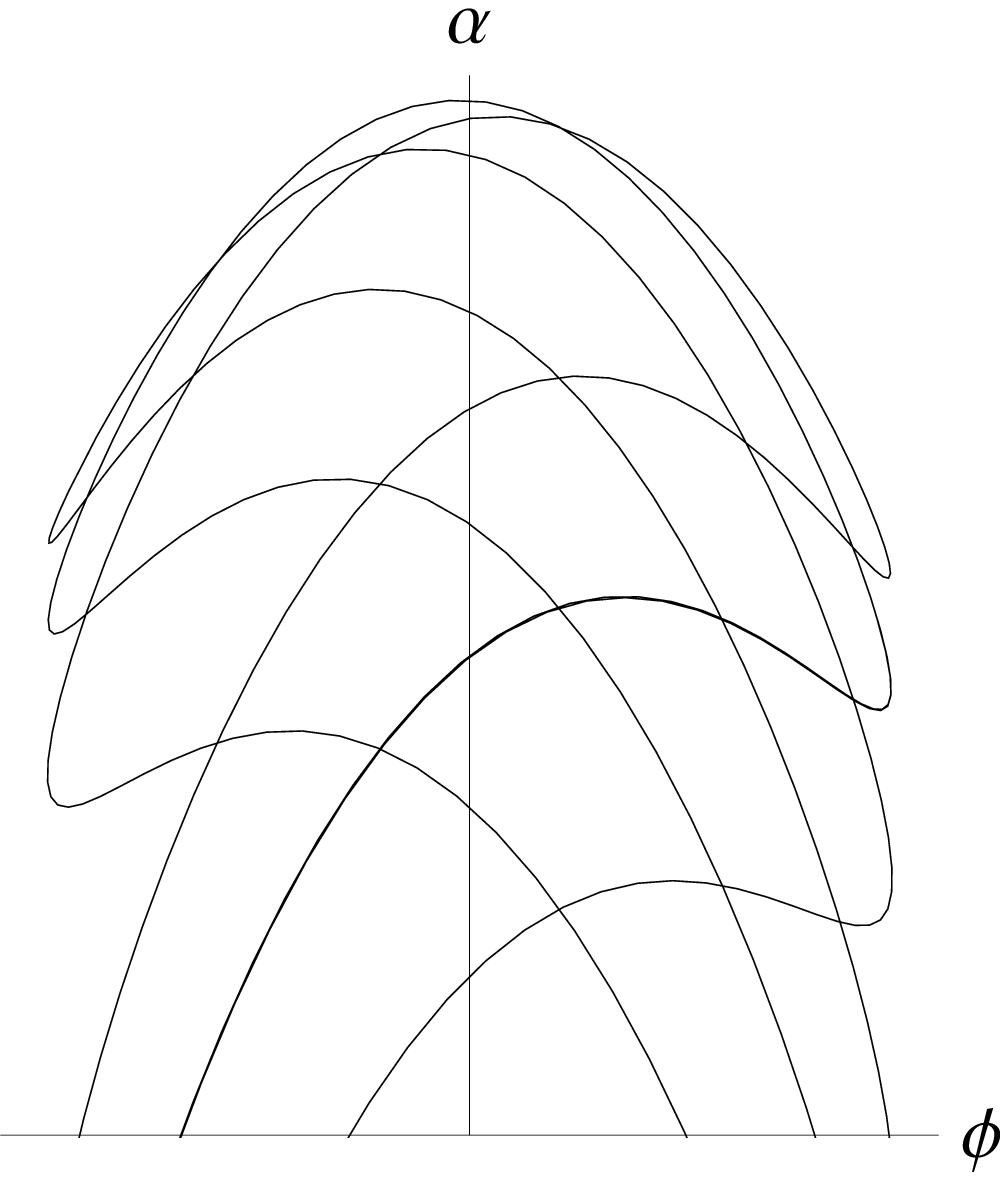}}\\
\subfigure[]{\label{fig:class_sol-c}\includegraphics[scale=.55]{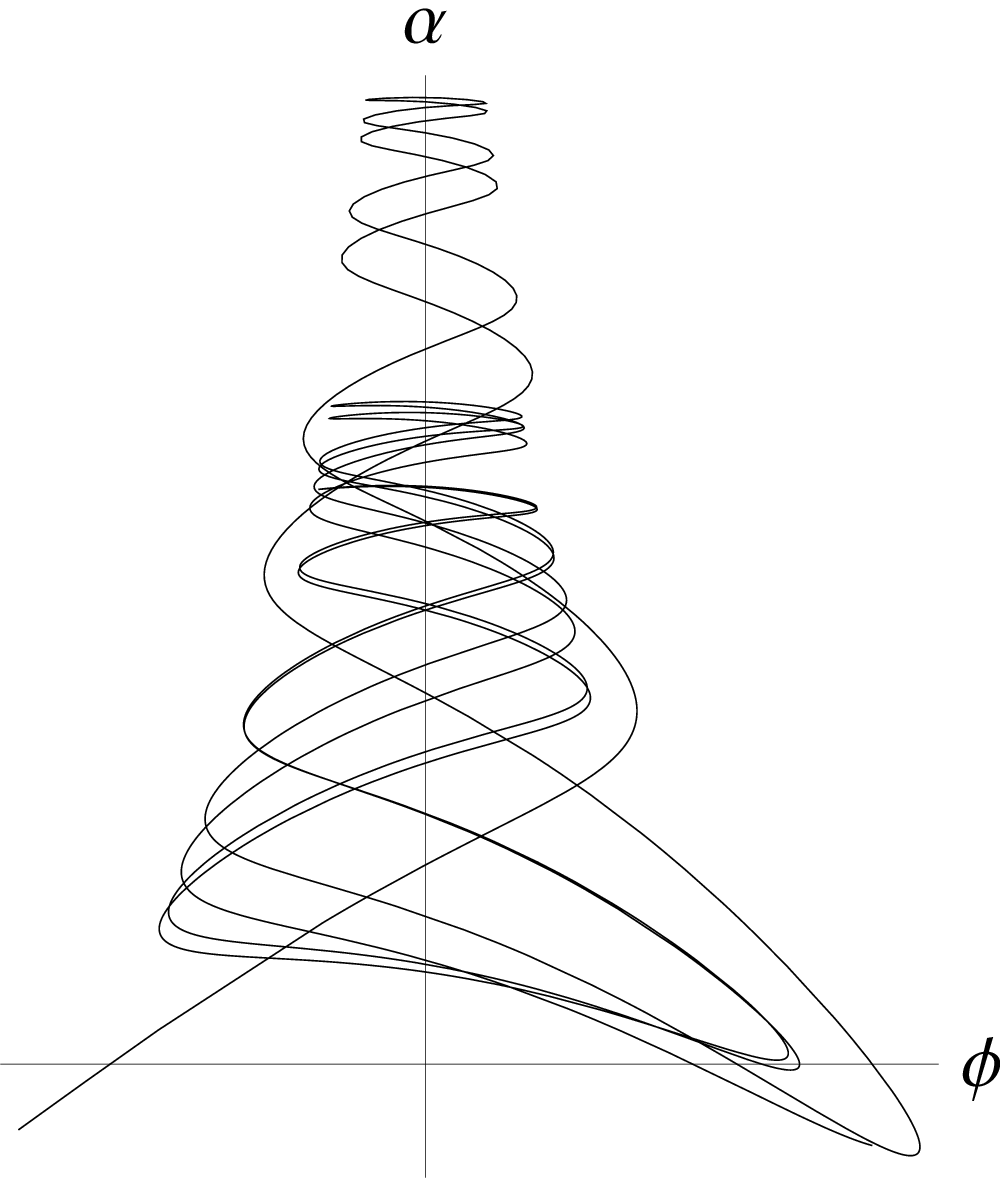}}\goodgap
    \subfigure[]{\label{fig:class_sol-d}\includegraphics[scale=.55]{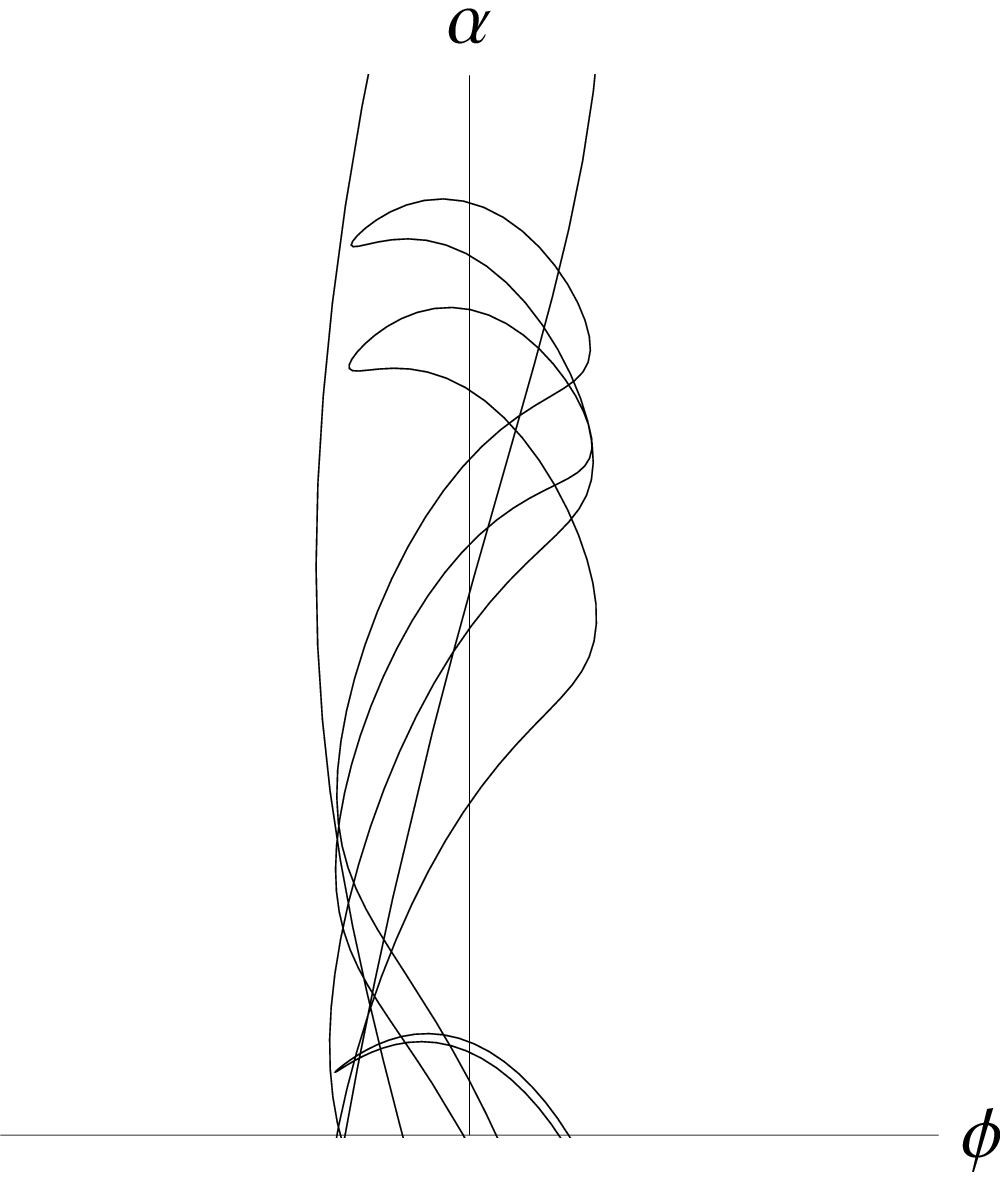}}
 \end{center}
\caption{\small Two typical classical solutions to the closed FRW spacetime---both $\phi$ and $a$ generically fail to be globally valid internal clock functions in this model. Here we used $\alpha=\ln(a)$ as appropriate for the canonical discussion following (\ref{hamcon2}, \ref{ham2}). Insets (a) and (c) show extended segments of (both the expanding and recontracting branches of) relational evolution up to the point of maximal expansion, $\alpha_{max}=\ln(a_{max})$. The (new) scale factor $\alpha$ oscillates between points of regular (nonglobal) maxima $\alpha_{max,k}=\ln(a_{max,k})$ and (nonglobal) minima $\alpha_{min,k}=\ln(a_{min,k})$; Inset (b) shows
a close--up of the same configuration space trajectory as (a) near
$\alpha_{max}$,
displaying the nonglobal extrema in a greater detail, while inset (d)
depicts a close--up on an intermediate section of the trajectory in
(c).}
  \label{fig:class_sol}
\end{figure*}

In fact, the situation for relational evolution appears even worse:
as noted in \cite{page,hawk1,hawk2,belinsky,cornshell}, there exists
a countably infinite discrete set of periodic solutions which bounce
without ever encountering a spacetime singularity. In
\cite{page,kkt,cornshell}, furthermore, it was shown that even an
uncountably infinite discrete set of perpetually bouncing aperiodic
solutions (of measure zero in the space of solutions
\cite{starobinsky,cornshell}) exists which exhibits an interesting
fractallike behavior. The system of (\ref{fried}, \ref{KG}) is thus
nonintegrable and chaotic \cite{page,kkt,cornshell}; this feature
lies at the root of many troubles in the quantum theory.

The reason for the absence of a globally valid internal clock
function in this model universe can be seen especially nicely in the
Hamiltonian formulation, which is required anyway in order to compare
with the effective results in Sec.~\ref{sec_frweff} below. For
practical purposes, let us perform a variable transformation
$\alpha= \ln(a)$ and henceforth work with $\alpha$. This is
convenient as, first, in the quantum theory one thereby avoids a
factor ordering problem in the Hamiltonian constraint
\cite{isham1,Kiefer} (see (\ref{wdw1}) below); second, the resulting
quantum Hamiltonian constraint (\ref{wdw1}) is explicitly of the
form of (\ref{qcon}), and thus the effective constructions of
Sec.~\ref{sec_semeff} are directly applicable; and third, we now
have $-\infty<\alpha<\infty$, and $-\infty<\phi<\infty$ and thus a
configuration space $\cq=\mathbb{R}^2$ which is somewhat simpler to
quantize than $\cq=\mathbb{R}\times\mathbb{R}_+$
\cite{isham1,isham2}.\footnote{For instance, $\hat{p}_a$ is {\it
not} self--adjoint on $L^2(\mathbb{R}_+,da)$. Or, when choosing
$L^2(\mathbb{R},da)$ instead, one would somehow have to give meaning
to $a<0$. On the other hand, $\hat{p}_\alpha$ {\it is} self--adjoint
on $L^2(\mathbb{R},d\alpha)$ and $-\infty<\alpha<+\infty$.} The big
bang and big crunch singularities will now appear at
$\alpha\rightarrow -\infty$, which is not an issue for our purposes,
since in the effective approach we shall be focusing on the regime
of maximal expansion of the scale factor $a$ (presumably, only a
full quantization can cope with the classically singular regime;
however, see \cite{singh}). For completeness, note, furthermore,
that when discussing the quantum dynamics in Secs.~\ref{sec_qt}
and~\ref{sec_frweff} below, small (big) fluctuations in $\alpha$ do
not necessarily translate into small (big) fluctuations in $a$.

\begin{figure}[htp]
  \begin{center}
    \subfigure[$\alpha(t)$ thick plot, $\phi(t)$ dashed plot]{\label{fig:class_can_var-a}\includegraphics[height=.3\textwidth,width=.35\textwidth]{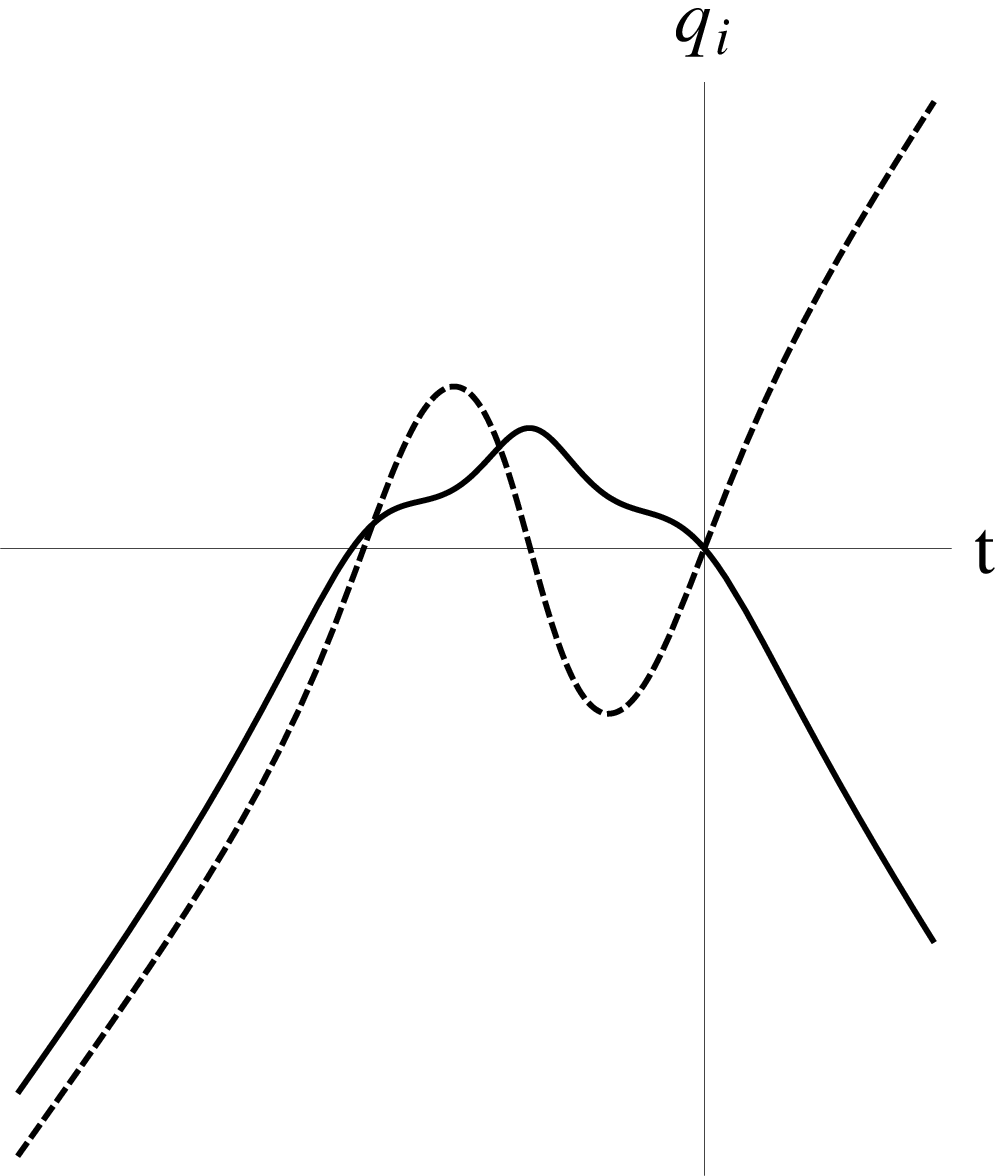}}
  \hspace{2cm}  \subfigure[$p_{\alpha}(t)$ thick plot, $p_{\phi}(t)$ dashed plot]{\label{fig:class_can_var-b}\includegraphics[height=.3\textwidth,width=.35\textwidth]{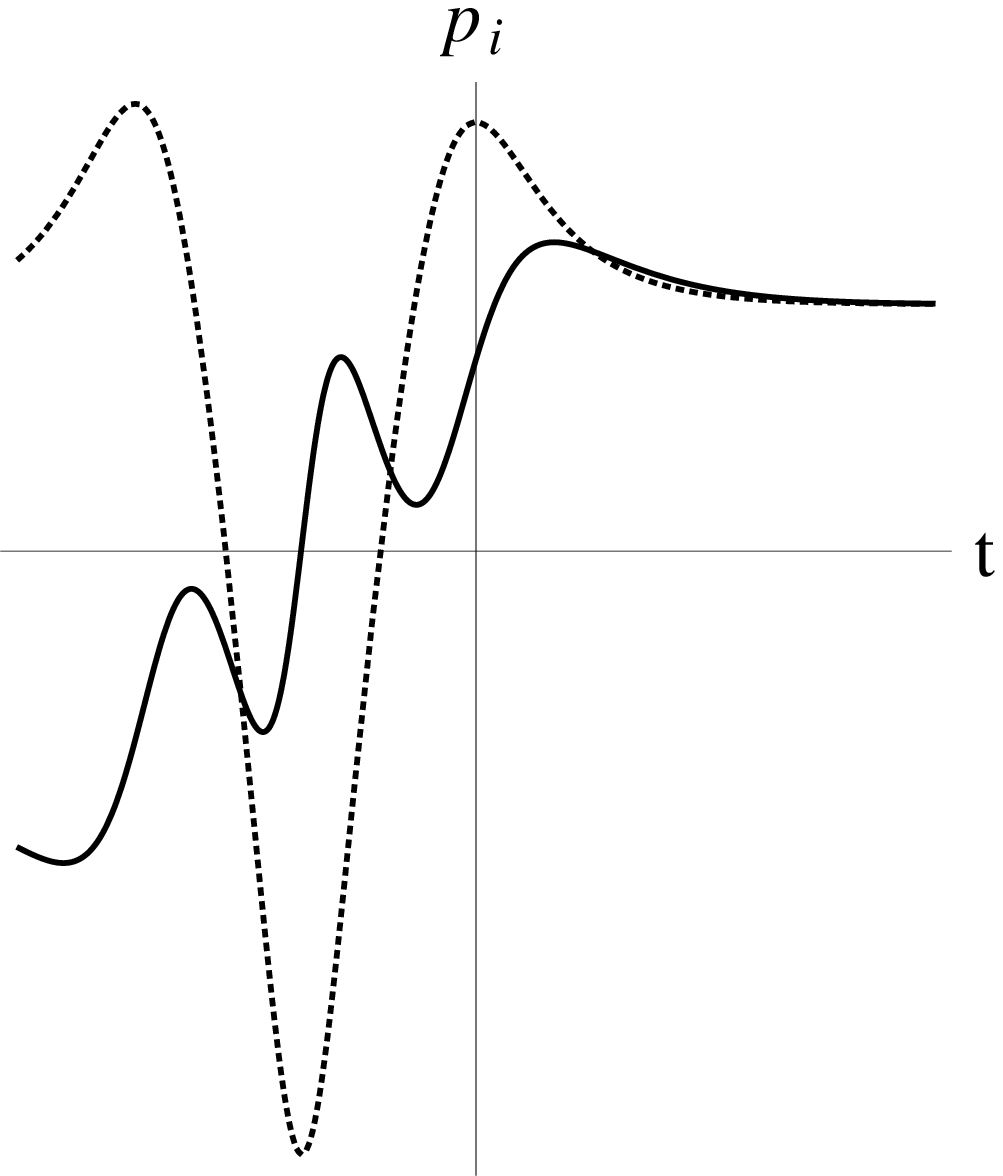}}
  \caption{\small Evolution of the canonical variables governed by (\ref{ham2}) for a rather benign classical solution. Notice how $\alpha$ features quasi--turning points close to the turning points of $\phi$ (also manifested in $p_\alpha$ having a local minimum close to the zeros of $p_\phi$). }
  \label{fig:class_can_var}
  \end{center}
\end{figure}

Choosing a gauge $N=e^{3\alpha}$, it is straightforward to
arrive at the expression for
the Hamiltonian constraint corresponding to the system
(\ref{action})
\cite{Kiefer,hajicek}:
\begin{equation}\label{hamcon2}
C_H=p_\phi^2-p_\alpha^2-e^{4\alpha}+m^2\phi^2e^{6\alpha}=0\q,
\end{equation} which is precisely of the form of (\ref{ccon}). 
The term
$m^2\phi^2e^{6\alpha}$ provides the coupling between the relational
clock, i.e.\ either $\alpha$ or $\phi$, and the evolving
configuration variable, i.e.\ either $\phi$ or $\alpha$,
respectively. In fact, the squared mass $m^2$ can be interpreted as
the coupling constant, while the factor $e^{6\alpha}$ can in certain
regimes be treated as an adiabatic factor \cite{marolf1,Kiefer}.
This coupling term will have a great effect on quantum relational
evolution. Using the symplectic structure on $T^*\cq$, the
corresponding canonical equations of motion read \ba\label{ham2}
\dot{\alpha}&=&\{\alpha,C_H\}=-2p_\alpha\q,\nn\\
\dot{p}_\alpha&=&\{p_\alpha,C_H\}=4e^{4\alpha}-6m^2\phi^2e^{6\alpha}\q,\nn\\
\dot{\phi}&=&\{\phi,C_H\}=2p_\phi\q,\nn\\
\dot{p}_\phi&=&\{p_\phi,C_H\}=-2m^2\phi e^{6\alpha}\q, \ea where now
the overdot refers to differentiation with respect to the coordinate
time $t$. As a consequence of $N=e^{3\alpha}$, note that henceforth
$t$ does {\it not} coincide with the proper times $\tau$ of comoving
observers in  (\ref{metric}). Fig.~\ref{fig:class_can_var} depicts
the behavior of the canonical variables for a rather benign
solution.

In a work concerning the precise origin of nonunitary relational evolution in the quantum theory of finite--dimensional parametrized systems \cite{hajicek}, H\'aj\'{\i}\v{c}ek has shown that unitarity requires the existence of a (temporally) global internal clock function already at the classical level which, in turn, was shown to be equivalent to the classical system being reducible. As an example, the system governed by (\ref{action}, \ref{hamcon2}) was considered, and it was shown that \cite{hajicek}
\begin{enumerate}
\item the constraint surface $\cc$ defined by (\ref{hamcon2}) in $T^*\cq$ is of topology $\cc=\mathbb{R}^2\times\mathbb{S}^1$ and is thus connected but not simply connected, and
\item the flow of $C_H$ on $\cc$ has no critical points, but incontractible cycles (around $\mathbb{S}^1$).
\end{enumerate}
The incontractible cycles, of course, correspond to the periodically bouncing solutions \cite{hawk1,page,kkt,cornshell} alluded to above. These cycles of the Hamiltonian flow on $\cc$ prevent the system from being (globally) reducible and possessing a global clock \cite{hajicek}.

\subsubsection{Classical relational dynamics and nonintegrability}\label{sec_nonin}

Let us make a few statements regarding relational evolution in this
nonintegrable model universe. We will not show here that the model is nonintegrable and chaotic since this has been demonstrated elsewhere \cite{page,kkt,cornshell}. We only summarize the facts relevant for our subsequent discussion. This discussion is of relevance, because in the
majority of the literature on relational dynamics, the possibility
of nonintegrability, despite it being a typical property of generic dynamical systems \cite{chaos} and having severe repercussions for relational evolution~\cite{smolin},
is largely ignored. We therefore believe that the results of the
present article are a first step towards a more general discussion
of the fate of relational dynamics, specifically in the quantum
theory. In particular, nonintegrability means that the system does
not possess any global constants of motion (i.e.\ Dirac observables)
other than the Hamiltonian itself \cite{chaos}.\footnote{In fact, in
the present model the Hamiltonian constraint (\ref{hamcon2})
coincides with the first integral of motion defined by the
Friedman equation (\ref{fried}).} Nevertheless, relational
evolution and Dirac observables may still exist {\it locally} (in
`time'), or rather implicitly and by means of the implicit function
theorem one could, in principle, still explicitly derive locally
valid observables.\footnote{For instance, in Eq.\ (5.6) of
\cite{Kiefer} the relational observable $\phi(a)$ is given for the
matter--dominated phase of expansion where $a\propto\tau^{2/3}$ and
$\tau$ is proper time.} This, certainly, features in the quantum
theory, and this is where we expect the effective approach to
relational evolution \cite{bht1,bht2} to come in handy, as it enables
one to make sense of local time evolution and (temporally) local
relational observables (aka {\it fashionables} in the terminology of
\cite{bht1,bht2}) in the semiclassical regime. However, even if
locally a complete set of relational observables is derived---in
contrast to integrable systems---this set in general no longer
characterizes the orbit, because chaotic systems typically possess
ergodic orbits which come arbitrarily close to any point on the
energy surface (i.e., for constrained systems, the constraint
surface) \cite{chaos}.

\begin{figure}[htp]
  \begin{center}
\includegraphics[scale=.65]{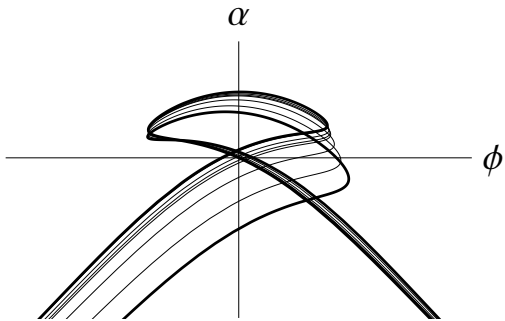}
  \end{center}
  \caption{\small Defocusing of nearby trajectories, caustics develop along the extrema of $\phi$ (see also \cite{page2}).}
  \label{fig:traj_chaos}
\end{figure}

Another generic---and related---property of chaotic systems is the
instability of initial data \cite{chaos}: chaotic systems generally
contain closed (periodic or unperiodic) orbits which are unstable in
the sense that a trajectory based on initial data arbitrarily close
to such a closed orbit will typically exponentially diverge from the
closed orbit and eventually become entirely
uncorrelated. Clearly, exponential divergence depends strongly on
the time coordinate, which is potentially dangerous in General
Relativity; however, there exists a very general definition of
chaotic behavior which takes this into account and essentially
requires a defocusing of trajectories (i.e., no statement is made
about how rapidly the defocusing occurs), as well as ergodicity
\cite{dyncosmo}. A further coordinate--independent measure of chaos, suitable for General Relativity, is the `topological entropy' which determines the complexity of the system by means of the set of closed orbits \cite{chaos}. The closed orbits of the present model universe
were described in detail in \cite{page,kkt,cornshell}. In
particular, in \cite{cornshell} the resulting fractal structure in
the space of initial data and the topological entropy were nicely exhibited, demonstrating how
solutions initially arbitrarily close can experience completely
unrelated fates.\footnote{It should be noted that this fractal structure is independent of the canonical variables chosen: tracing an unstable closed orbit requires infinite fine--tuning of the canonical initial variables. This will hold in any set of canonical variables related by nonsingular transformations \cite{chaos}.} In fact, defocusing of nearby trajectories occurs
in the present model already for trajectories not arbitrarily close to
a closed orbit, albeit in a much milder fashion. For instance, Fig.~\ref{fig:traj_chaos}
depicts how neighboring trajectories fan out in the region of maximal expansion already for a rather well--behaved classical solution. For generic solutions exhibiting more oscillations in both $\phi$ and $\alpha$ \cite{kkt}, this feature will get more pronounced. Such defocusing will be particularly relevant in the quantum theory, since it constitutes the ultimate cause of a generic breakdown of semiclassicality and relational evolution.


Finally, classically there is no obstruction to using either $\alpha$ or $\phi$ as a global clock function, despite the turning points of the clock variables and the ensuing multivaluedness of the relational observables, because we can always resort to the gauge parameter in order to provide an ordering to the correlations \cite{bht2}. Nevertheless, it is more practical to employ $\alpha$ as an internal clock for large parts of the evolution due to the highly oscillatory nature of the scalar field at large volumes. In the quantum theory, it will no longer be possible to employ either variable globally due to nonunitarity and a breakdown of evolution {\it before} classical turning points.

Classically, one imposes suitable (compatible with $C_H$) initial
data at some fixed $t=t_0$, which can be translated into a
relational initial value problem (IVP): when using $\alpha$ as the clock, one could choose $\phi(\alpha_0)$ and $p_\phi(\alpha_0)$ at some value $\alpha_0=\alpha(t_0)$, which, e.g., corresponds to some configuration on the expanding branch of cosmic evolution, if one chooses the (here due to (\ref{ham2})) negative sign solution for the initial clock momentum $p_\alpha(t_0)$ via the constraint of (\ref{hamcon2}).
Indeed, in relativistic systems subject to constraints quadratic in
the momenta, a relational IVP additionally requires an initial
internal time direction (i.e.\ a sign for the clock momentum) in order to relationally evolve
\cite{bht2,haj2}. This initial data is subsequently evolved through the
maximal extension into the big crunch singularity, such that the
contracting branch is classically the logical successor of the
expanding branch. In contrast to earlier work
\cite{zeh,Kiefer,zehbook} on the quantum theory of (\ref{action}),
we shall perform the same IVP construction for sufficiently
semiclassical states in the effective framework in
Sec.~\ref{sec_frweff} below.

\subsection{Troubles for Hilbert space quantizations}\label{sec_qt} 

The classical nonintegrability of the model suggests a rather
complicated quantum dynamics. Indeed, generally the transition from
quantum to classical is a highly nontrivial challenge in chaotic
models and qualitatively quite distinct from the analogous task for
nonchaotic systems \cite{chaos}. While substantial research has
been devoted to gaining a general (but mostly approximate)
understanding of at least the semiclassical solutions to the present
model in various approaches
\cite{hawk1,hawk2,page2,Kiefer,zehbook,singh,parker}, dynamical
(relational) questions have thus far not been properly addressed.
This is simply because no (nontrivial) exact quantum solutions are
known, let alone a physical inner product on the space of solutions
in which one could compute expectation values of various quantities.
In order to be able to compare with the effective relational
dynamics of Sec.~\ref{sec_frweff}, we ideally would like to extract
(at least approximate) dynamical information from the Hilbert space
or path integral quantizations carried out thus far. In the present
section, we wish to explain why it is practically difficult to
extract relational dynamics from any of the previous approaches.

To this end, we firstly recall a result presented in \cite{bht2}:
the effective relational dynamics in a toy model devoid of global
clocks was shown to be equivalent at order $\hbar$ to the dynamics
(of expectation values) in a (temporally) local internal time
Schr\"odinger regime. The latter is obtained by a local
deparametrization of the classical model and a subsequent
quantization. For a model governed by a quadratic constraint of the
form $C=p_t^2-H^2(t,q,p)$, one can locally deparametrize by choosing
a local clock, say, $t$ and factorizing the constraint as
$C=C_+C_-=(p_t+H)(p_t-H)$. Standard quantization of $C_\pm$ yields a
Schr\"odinger equation with `time--dependent' square--root
Hamiltonian $\hat{H}$ (defined by spectral decomposition):
\ba\label{schrod} i\hbar\partial_t\psi(t,q)=\pm
\hat{H}(t,\hat{q},\hat{p})\psi(t,q), \ea which, if $t$ is a nonglobal
internal clock, is only locally valid (in `time' $t$) because of
nonunitarity. One may thus wonder whether a similar construction
could be performed for the present model universe such that we may
compare the local dynamics of the Schr\"odinger regime with the
effective results. In fact, this question was already considered
(for very different reasons) in an early work on quantum cosmology
by Blyth and Isham \cite{isham}, in which they investigated a
reduced quantization of FRW models filled with a homogenous scalar
field. They considered various choices of relational time variables
(chosen before quantization) which all yield distinct
time--dependent Schr\"odinger equations with square--root
Hamiltonians that describe precisely the desired Schr\"odinger
regimes.\footnote{One of the motivations of \cite{isham} to quantize
by the reduction procedure, instead of a Dirac quantization leading
to a Wheeler--DeWitt equation was to avoid the nonpositive
definiteness of Klein--Gordon--type inner products.} As regards the
relation between the Schr\"odinger regime and a Dirac quantization
yielding a Wheeler--DeWitt (WDW) equation (with quantized
$\hat{t}$), \begin{eqnarray}\label{WDW}
\hat{H}^2(\hat{t},\hat{q},\hat{p})\tilde{\psi}(q,t)=\hat{p}_t^2\tilde{\psi}(q,t)=-\hbar^2\partial^2_t\tilde{\psi}(q,t),
\end{eqnarray}
it was noted in \cite{isham} that (\ref{WDW}) does {\it not} follow
from (\ref{schrod}) when $\hat{H}$ is explicitly time dependent,
because acting with $\pm\hat{H}$ on both sides of
(\ref{schrod})---rather than (\ref{WDW})---yields
\ba
\hat{H}^2(t,\hat{q},\hat{p})\psi(t,q)=-\left(\hbar^2\partial^2_t
 \pm
i\hbar\partial_t\hat{H}(t,\hat{q},\hat{p})\right)\psi(t,q).\nn
\ea
However, in \cite{bht1} it was shown that to {order $\hbar$, the
expectation value versions\footnote{Assuming a standard $t=const$
Schr\"odinger theory inner product away from any turning points.} of
(\ref{schrod}) and (\ref{WDW}) {\it are}, in fact, solved by the
{\it same} state $\psi=\tilde{\psi}$ if the expectation value of the
`internal time operator' $\hat{t}$ in (\ref{WDW}) is complex with
an imaginary part coinciding with the effective result (\ref{imt}) \ba
\Im[\langle\hat{t}\rangle]=-\frac{\hbar}{2\langle\hat{p}_t\rangle}.
\ea To semiclassical order, the Schr\"odinger regime may thus be
understood as locally approximating a solution to the relativistic
WDW equation (away from classical turning points).

However, the explicit construction of solutions in \cite{isham} was
only carried out for the $k=1$, $m=0$ and $k\leq0$, $m\neq0$
FRW models. The reason for avoiding the present model is explained
as follows: in our case (\ref{hamcon2}), the classical Hamiltonian
for evolution in $t=\alpha$ time is given by
$H(\alpha;\phi,p_\phi)=\sqrt{p_\phi^2-e^{4\alpha}+m^2\phi^2e^{6\alpha}}$,
while the one for evolution in $t=\phi$ time reads
$H(\phi;\alpha,p_\alpha)=\sqrt{p_\alpha^2+e^{4\alpha}-m^2\phi^2e^{6\alpha}}$.
The ensuing quantum Hamiltonian $\hat{H}(t,\ldots)$ is not only
`time--dependent;' it also fails to commute with itself at different
`times', $[\hat{H}(t,\ldots),\hat{H}(t',\ldots)]\neq0$ for both
$t=\alpha,\phi$. Consequently, `energy' eigenstates at a given
`time' fail to be eigenstates at later `times' and the formal
solution to (\ref{schrod}) involves a Dyson time ordering \ba
\psi(t,q)=\hat{U}(t,t_0)\psi(t_0,q)
=T\left[\exp\left(\mp\frac{i}{\hbar} \int^t_{t_0}
\hat{H}(s,\hat{q}, \hat{p})ds\right)\right]\psi(t_0,q). \nn \ea
Constructing explicitly the time evolution operator
$\hat{U}(t,t_0)$ with either
$\hat{H}(\alpha;\hat{\phi},\hat{p}_\phi)$ or
$\hat{H}(\phi;\hat{\alpha},\hat{p}_\alpha)$, unfortunately, does not
(even to order $\hbar$) seem feasible for this nonintegrable
system. We thus abstain from further attempting to construct a local
Schr\"odinger regime.

Next, in order to extract relational dynamics from the quantum
theory, one could try to solve the WDW equation and consider a
suitable inner product in order to compute expectation values which
may be compared to the effective results. The canonical Dirac
quantization was considered, e.g., in \cite{page2,Kiefer,zehbook}.
The standard quantization of (\ref{hamcon2}) yields a Klein--Gordon--type hyperbolic partial differential equation (setting for now
$\hbar=1$),\footnote{Note that the choice of variables and (in this
case, trivial) factor ordering here is such that the derivative terms
constitute the invariant d'Alembertian with respect to the
minisuperspace metric (\ref{smet}).} \ba\label{wdw1}
\left(\frac{\partial^2}{\partial\alpha^2}-\frac{\partial^2}{\partial\phi^2}-e^{4\alpha}+m^2\phi^2e^{6\alpha}\right)\psi(\alpha,\phi)=0,
\ea with variable mass $M^2=e^{4\alpha}(e^{2\alpha}m^2\phi^2-1)$ in
the two--dimensional Lorentzian superspace metric \ba\label{smet}
ds^2=-d\alpha^2+d\phi^2. \ea Thus, $\alpha=const$ is a spacelike
slice in minisuperspace.

WKB approximations to this equation have been extensively studied,
e.g., in \cite{Kiefer,hawk1,page2,cornshell,lafshell}, from various
perspectives, all reporting a breakdown of this semiclassical
expansion in the region of maximal extension. A WKB approximation
$\psi=\sum_nC_n(\alpha,\phi)\exp(\pm iS_n(\alpha,\phi))$ is valid
only if the amplitude $C_n$ varies much more slowly than the phase $S_n$
\cite{hawpage,lafshell,hawk1,Kiefer,page2}. As pointed out in
\cite{cornshell,page2}, the caustics resulting from focusing of
nearby classical trajectories (see also Fig.
\ref{fig:traj_chaos} above) cause $|C_n|^2\rightarrow\infty$, while
$|C_n|^2$ goes rapidly to zero, where classical trajectories defocus,
for instance, in the region of maximal expansion (also in
Fig.~\ref{fig:traj_chaos}),
which leads to a generic breakdown of the WKB approximation. This is
of relevance for at least a qualitative comparison to the effective
results displayed in Sec.~\ref{sec_frweff} below. Consequently, we
wish to summarize the pivotal features of previous semiclassical
constructions.

For instance, Kiefer \cite{Kiefer} imposed initial data for $\psi$ on a spacelike slice $\alpha=const$ in order to construct wave packets in minisuperspace, approximately solving (\ref{wdw1}) via a Born--Oppenheimer (with expansion parameter $m_p^{-1}$) and a subsequent WKB approximation. Tubelike standing waves representing classically expanding and contracting universes could be constructed if an additional `final condition' in $\alpha$, namely $\psi\rightarrow0$ as $\alpha\rightarrow\infty$, was imposed for reasons of `normalizability'.\footnote{While sensible in the construction of \cite{Kiefer}, the `final condition' should not be viewed as a normalization condition, because normalization requires an inner product. In fact, in \cite{page2} no `final condition' was imposed and the wave function was not strongly damped for large $\alpha$, which was interpreted as leading to a high probability that the universe would be large compared to the Planck length. (Although a probabilistic interpretation, again, requires a consistent inner product which, as discussed below, awaits identification.)}  The turning point $\alpha_{max}(n)$ of the individual oscillator modes in the wave packet depends strongly on the mode $n$, and thus the reflection of the wave packet at the average $\alpha_{max}=\alpha_{max}(\bar{n})$ is described by a (chaotic) scattering phase shift which depends on the mass and is a multiple of $\pi$ only for discrete values of $m$ \cite{Kiefer}. Narrow wave tubes on both the expanding and recontracting branch can thus only be constructed for these special values of $m$ and only away from the classical turning region, i.e.\ only for $\alpha<<\alpha_{max}$.
Furthermore, Hawking applied the ``no boundary'' proposal \cite{hh} (which renders an IVP superfluous) to the present model \cite{hawk1}.
The ensuing (semiclassical) wave function can be interpreted as a superposition of quantum states peaked around an ensemble of nonsingular bouncing solutions with long inflationary period which correspond to the aforementioned set of measure zero periodic and aperiodic solutions \cite{page,kkt,cornshell}.\footnote{This is in agreement with standard results on the semiclassical limit of quantum models which are classically chaotic. Semiclassical states are typically concentrated on the closed orbits of measure zero \cite{chaos}.} Numerical evidence for these results was exhibited in \cite{hawk2}, while similar outcomes with special attention to singular classical trajectories were reported in \cite{lafshell}. Page \cite{page2} approximated the Hawking wave function by starting from the canonical constraint (\ref{wdw1}) and translating the ``no boundary'' condition into sufficient Cauchy data. Also, this WKB approximation breaks down due to caustics at the extrema of $\phi$ \cite{page2}.

As regards the classical determinism, mentioned in
Sec.~\ref{sec_nonin}, of having the recontracting branch as the
logical successor of the expanding one, it was maintained in
\cite{zeh,Kiefer,zehbook} that:
\begin{itemize}
\item[(1)] The quantum IVP is very different from that of the classical theory. Initial data has to be imposed on {\it all} of the minisuperspace slice $\alpha=const$, implying that both branches have to be there `initially' (in $\alpha$). `Initial' and `final state' can no longer be distinguished.
\item[(2)] It is meaningless in quantum cosmology to extend classical paths through the turning region of $\alpha$ into the recollapsing phase. The WKB approximation does not provide the complete classical trajectory. The latter could only be obtained through continuous measurement by higher degrees of freedom (which would suppress the scattering at $\alpha_{max}$).
\end{itemize}
However, these statements partially depend on the construction used in \cite{Kiefer}, namely, (a), on obtaining the semiclassical limit by means of a WKB approximation, (b), on using solely $\alpha$, rather than $\phi$, as the internal clock and, (c), on the `final condition', $\psi\rightarrow0$ as $\alpha\rightarrow\infty$. Let us discuss these one by one.

(a) While a WKB approximation is one way of obtaining semiclassical
information from a quantum model, it is not the most general
semiclassical approximation and necessarily breaks down for chaotic
systems \cite{cornshell}. On the other hand, the semiclassical
approximation employed in the effective approach is very general in
nature. We shall see in Sec.~\ref{sec_frweff} that semiclassicality
{\it can} be achieved in the classical turning region; however, only
for sufficiently peaked initial effective states. A fairly classical
trajectory with the recollapsing branch being the logical
successor of the expanding one can thus be obtained {\it without}
decoherence of additional degrees of freedom.

(b) The (chaotic) scattering of the wave packet around
$\alpha_{max}$ \cite{Kiefer,zehbook} manifests nonunitarity of evolution in the nonglobal clock
$\alpha$. Indeed, as pointed out in \cite{bht1}, the
interference of segments of the wave function/packet before and
after the turning region of a nonglobal internal time
function---as a result of different modes having different turning
points---leads to a superposition of (internal) time directions,
i.e.\ of positive and negative frequencies associated to the
spectrum of the momentum conjugate to the internal clock function.
This necessarily leads to a breakdown of the evolution in the nonglobal clock and of any deparametrization---aimed at approximating the physical state and employing inner products based on the level surfaces of the clock---{\it before} the classical turning point. This is in agreement with
the analysis in \cite{marolf1} concerning the reconstruction of the
unitary Schr\"odinger and Heisenberg picture from relational quantum
dynamics, which turns out to be only locally feasible for
sufficiently semiclassical states and clock degrees of freedom far
enough away from any clock turning points. Unitarity of evolution in a given time
variable in a deparametrization is tantamount to the preservation of inner products in the
evolution, which is evidently not
possible here with inner products based on the level surfaces of the nonglobal clock. Instead, one could switch to relational evolution in
a new clock if it behaves sufficiently semiclassically in the
turning region of the first clock \cite{bht1,bht2}. If no such
degree of freedom is admitted by the state, relational evolution
necessarily breaks down. In this case the general quantum state may, indeed, no longer be interpreted and approximated dynamically by means of a deparametrization, and rather the full WDW equation and the proper {\it physical} inner product are required in general, which may not admit a global relational interpretation. In the present model universe, however, $\phi$ may be used for sufficiently benign and semiclassical states as an intermediate clock in the turning region of $\alpha$. Whereas it is not clear how this could be achieved at the level of the WDW equation (for which one would also need a more general semiclassical solution than obtained by means of a WKB approximation), this is precisely what will be carried out in the effective framework in Sec.~\ref{sec_frweff}. At the effective level, the nonlocal initial value problem and single evolution generator of \cite{zeh,Kiefer,zehbook} is traded for a local initial value problem solely imposed, say, on the expanding branch, and for the necessity of two evolution generators, one in $\alpha$, the other in $\phi$ time. In this manner, but only for sufficiently semiclassical states, `initial' and `final' states {\it can} be distinguished and connected by (semi--)classical paths in the turning region.

(c) In fact, it is the `final condition' which prevents narrow wave packets around $\alpha_{max}$; only exponentially (in $\alpha$) decreasing modes are allowed, and the data for both the expanding and recontracting branch must be present initially at $\alpha_0$, but is subsequently scattered at $\alpha_{max}$ \cite{Kiefer,zehbook,zeh}. On the other hand, no final condition can be imposed in the effective approach which for sufficiently benign states, nonetheless, yields semiclassical trajectories in the region of maximal expansion.

Let us consider (na\"ive) possibilities for an inner product. (i) Since the operator $\hat{H}^2=-\partial^2_\phi-e^{4\alpha}+m^2\phi^2e^{6\alpha}$ is not generally nonnegative, evolution with respect to $\alpha$ is nonunitary, and a standard Schr\"odinger--type inner product is clearly not preserved. (ii) Group averaging \cite{groupav} is commonly employed in constructing physical inner products in quantum cosmology; however, it requires integrating over the flow of the quantum constraint, which does not seem practical on account of the classical nonintegrability. (iii) There exists a method going back to DeWitt \cite{haj2,DW} which yields a conserved quadratic form on $\ch_{phys}$ from $\ch_{aux}$ which in the present case is just $L^2(\mathbb{R}^2,d\alpha d\phi)$:
\begin{Theorem} Let $(\cq,\eta)$ be an n-dimensional configuration manifold with volume form $\eta$, and $\hat{C}$ be a second--order differential operator on $\cc^2_0(\cq,\mathbb{C})$ (space of twice--differentiable complex functions with compact support on $\cq$) that is symmetric with respect to the scalar product on $L^2(\cq,\eta)$. Then, for any $\Psi,\Phi\in\cc^2_0(\cq,\mathbb{C})$, there is a vector field $\vec{J}[\Psi,\Phi]$ on $\cq$ such that
\ba
(\hat{C}\Psi)^*\Phi-\Psi^*(\hat{C}\Phi)=\text{Div}_\eta\vec{J}.
\ea
\end{Theorem}
Clearly, if both $\Psi$ and $\Phi$ are annihilated by a hyperbolic $\hat{C}$, $\vec{J}$ defines a conserved current on the space of solutions to $\hat{C}\psi=0$. It is not difficult to convince oneself that for the constraint of (\ref{wdw1}) $\vec{J}$ is just given by the standard Klein--Gordon current vector,
\ba
J^a=g^{ab}[(\partial_a\Psi^*)\Phi-\Psi^*(\partial_a\Phi)],
\ea
where $g^{ab}$ is the inverse two--dimensional minisuperspace metric (\ref{smet}), such that the conserved quadratic form provided by the theorem coincides with the Klein--Gordon inner product. Unlike in the case of a Klein--Gordon particle, we cannot restrict ourselves here globally to positive or negative frequency modes with respect to one of our clocks (on whose subspaces the Klein--Gordon inner product would be positive definite), because no global clock exists, and `positive' and `negative frequencies' in $\alpha$ time will necessarily mix up in the turning region of $\alpha$. In addition, the Klein--Gordon charge is identically zero for real $\Psi,\Phi$ and thereby trivially conserved. The semiclassical (approximate) solutions of \cite{hawk1,hawk2,page2,Kiefer} are real. Hence, it is not even possible to use the Klein--Gordon inner product as an approximation for known semiclassical states on only the `negative' (i.e., expanding) or `positive frequency' (i.e., recollapsing) branch away from the turning region in which frequencies mix up. It, therefore, remains unclear what the correct physical inner product should be
and how the {\it Hilbert space problem} could be solved.

In conclusion, relational dynamics of this nonintegrable model seems currently only practically feasible in the effective approach (and also there only in a limited regime), since it sidesteps many technical difficulties associated with a Hilbert space quantization \cite{bht2}.

\subsection{Effective relational dynamics}\label{sec_frweff}

Following the general procedure laid down in Secs.~\ref{sec_eff}
and~\ref{sec_semeff}, we now turn to the effective treatment of the
closed FRW model. Since we are only interested in the semiclassical
regime, we need not solve the full quantum dynamics but only `expand
around' classical trajectories. The nonintegrability is hence not
a {\it technical} problem for us when studying semiclassical states
corresponding to (nonclosed) classical solutions. We will study
rather benign trajectories; however, it will quickly become evident
what will happen for more generic and complicated solutions.

Using the potential $V(\alpha,
\phi)=e^{4\alpha}-m^2\phi^2e^{6\alpha}$ in (\ref{eq:deparam_effC}),
the constraint (\ref{wdw1}) translates to order $\hbar$ into the
following five quantum constraint functions:\footnote{The last two terms in the expression for $C$ are missing in the PRD published version of this article.}
\begin{align}\label{constraints_gen}
C&={p}_{\phi}^2+(\Delta p_{\phi})^2-{p}_{\alpha}^2-(\Delta p_{\alpha})^2-e^{4{\alpha}}-8e^{4{\alpha}}(\Delta \alpha)^2+m^2\phi^2e^{6{\alpha}}+m^2e^{6{\alpha}}(\Delta\phi)^2\nn\\
&\q+12m^2\phi
e^{6\alpha}\Delta(\alpha\phi) + 18m^2\phi^2e^{6\alpha}(\Delta\alpha)^2,\\
C_{\alpha}&=2p_{\phi}\Delta(\alpha p_{\phi})-2p_{\alpha}\Delta(\alpha p_{\alpha})-i\hbar\, p_{\alpha}+2m^2\phi e^{6\alpha}\Delta(\alpha\phi)+(6m^2\phi^2 e^{6\alpha}-4e^{4\alpha})(\Delta \alpha)^2,\nonumber\\
C_{\phi}&=2p_{\phi}\Delta(\phi p_{\phi})+i\hbar\, p_{\phi}-2p_{\alpha}\Delta(\phi p_{\alpha})+(6m^2\phi^2 e^{6\alpha}-4e^{4\alpha})\Delta(\alpha\phi)+2m^2\phi e^{6{\alpha}}(\Delta\phi)^2,\nonumber\\
C_{p_{\alpha}}&=2p_{\phi}\Delta(p_{\alpha}p_{\phi})-2p_{\alpha}(\Delta
p_{\alpha})^2 + (6m^2\phi^2 e^{6\alpha}-4e^{4\alpha})\Delta(\alpha
p_{\alpha}) + 2m^2\phi e^{6\alpha}\Delta(\phi p_{\alpha})
\nonumber \\ &\qquad-i\hbar\,(3m^2\phi^2 e^{6\alpha}-2\, e^{4\alpha}),\nonumber\\
C_{p_{\phi}}&=2p_{\phi}(\Delta
p_{\phi})^2-2p_{\alpha}\Delta(p_{\alpha}p_{\phi})+(6m^2\phi^2
e^{6\alpha}-4e^{4\alpha})\Delta(\alpha p_{\phi})+2m^2\phi
e^{6{\alpha}}\Delta(\phi p_{\phi})-i\hbar\, m^2\phi
e^{6\alpha}.\nonumber
\end{align}
Due to the degeneracy in the quantum Poisson structure, the five
constraints of (\ref{constraints_gen}) generate only four independent
gauge flows. To remove redundant degrees of freedom, we choose a
relational clock and a corresponding \emph{Zeitgeist}, thereby
fixing three of the four independent gauge flows. We are then left
with just one (Hamiltonian) constraint governing the evolution of
the system.

\subsubsection{Evolution in $\alpha$}\label{sec_evalpha}

Choosing $\alpha$ as our relational clock, we resort to the $\alpha$--\emph{Zeitgeist}
\begin{align}
(\Delta\alpha)^2=\Delta(\phi\alpha)=\Delta(\alpha p_{\phi})=0,
\end{align}
which, as can be easily checked by solving $C_\alpha$, leads to a saturation of the generalized uncertainty relation for the clock degrees of freedom.
The rest of the constraints is simplified as
\begin{align}\label{constraints_alpha}
C&={p}_{\phi}^2+(\Delta p_{\phi})^2-{p}_{\alpha}^2-(\Delta p_{\alpha})^2-e^{4\alpha}+m^2\phi^2e^{6\alpha}+m^2e^{6\alpha}(\Delta\phi)^2,\nonumber\\
C_{\phi}&=2p_{\phi}\Delta(\phi p_{\phi})+i\hbar\, p_{\phi}-2p_{\alpha}\Delta(\phi p_{\alpha})+2m^2\phi e^{6{\alpha}}(\Delta\phi)^2,\nonumber\\
C_{p_{\alpha}}&=2p_{\phi}\Delta(p_{\alpha}p_{\phi})-2p_{\alpha}(\Delta p_{\alpha})^2+2m^2\phi e^{6\alpha}\Delta(\phi p_{\alpha})-i\hbar(6m^2\phi^2 e^{6\alpha}-4e^{4\alpha}),\nonumber\\
C_{p_{\phi}}&=2p_{\phi}(\Delta p_{\phi})^2-2p_{\alpha}\Delta(p_{\alpha}p_{\phi})+2m^2\phi e^{6{\alpha}}\Delta(\phi p_{\phi})-i\hbar\, m^2\phi e^{6\alpha},
\end{align}
and can be used to solve for the unphysical moments $\Delta(\phi p_{\alpha}), (\Delta p_{\alpha})^2, \Delta(p_{\alpha}p_{\phi})$. Relational evolution of the remaining degrees of freedom in $\alpha$ is generated by the remaining first class (Hamiltonian) constraint which, by (\ref{eq:effCham_q1}) in the $\alpha$--\emph{Zeitgeist} reads
\begin{align}\label{Cham_gen}
C_H&={p}_{\phi}^2-{p}_{\alpha}^2-e^{4\alpha}+m^2\phi^2e^{6\alpha}+\left[1-\frac{p_{\phi}^2}{p_{\alpha}^2}\right](\Delta p_{\phi})^2-\frac{2m^2\phi e^{6{\alpha}}p_{\phi}}{p_{\alpha}^2}\Delta(\phi p_{\phi})\nonumber\\&\qquad+\left[m^2e^{6\alpha}-\frac{m^4\phi^2 e^{12{\alpha}}}{p_{\alpha}^2}\right](\Delta\phi)^2+i\hbar\,\frac{3m^2\phi^2 e^{6\alpha}-2e^{4\alpha}}{p_{\alpha}}.
\end{align}
Through the Poisson structure (\ref{poisson}), this constraint generates the following equations of motion:
\begin{align}\label{eom_alpha}
\dot{\alpha}&=-2p_{\alpha}+\frac{2p_{\phi}^2}{p_{\alpha}^3}(\Delta p_{\phi})^2+\frac{4m^2\phi e^{6{\alpha}}p_{\phi}}{p_{\alpha}^3}\Delta(\phi p_{\phi})+\frac{2m^4\phi^2 e^{12{\alpha}}}{p_{\alpha}^3}(\Delta\phi)^2-i\hbar\,\frac{3m^2\phi^2 e^{6\alpha}-2e^{4\alpha}}{p_{\alpha}^2},\nonumber\\
&\qquad\nonumber\\
\dot{p_{\alpha}}&=4e^{4\alpha}-6m^2\phi^2e^{6\alpha}+\frac{12m^2\phi e^{6{\alpha}}p_{\phi}}{p_{\alpha}^2}\Delta(\phi p_{\phi})-\left[6m^2e^{6\alpha}-\frac{12m^4\phi^2 e^{12{\alpha}}}{p_{\alpha}^2}\right](\Delta\phi)^2\nonumber\\&\qquad-i\hbar\,\frac{18m^2\phi^2 e^{6\alpha}-8e^{4\alpha}}{p_{\alpha}},\nonumber\\
&\qquad\nonumber\\
\dot{\phi}&=2p_{\phi}-\frac{2p_{\phi}}{p_{\alpha}^2}(\Delta p_{\phi})^2-\frac{2m^2\phi e^{6{\alpha}}}{p_{\alpha}^2}\Delta(\phi p_{\phi}),\\
&\qquad\nonumber\\
\dot{p_{\phi}}&=-2m^2\phi e^{6\alpha}+\frac{2m^2e^{6{\alpha}}p_{\phi}}{p_{\alpha}^2}\Delta(\phi p_{\phi})+\frac{2m^4\phi e^{12{\alpha}}}{p_{\alpha}^2}(\Delta\phi)^2-i\hbar\,\frac{6m^2\phi e^{6\alpha}}{p_{\alpha}},\nonumber\\
&\qquad\nonumber\\
\dot{(\Delta\phi)^2}&=4\left[1-\frac{p_{\phi}^2}{p_{\alpha}^2}\right]\Delta(\phi p_{\phi})-\frac{4m^2\phi e^{6{\alpha}}p_{\phi}}{p_{\alpha}^2}(\Delta\phi)^2,\nonumber
\end{align}
\begin{align}
\dot{\Delta(\phi p_{\phi})}&=2\left[1-\frac{p_{\phi}^2}{p_{\alpha}^2}\right](\Delta p_{\phi})^2+2\left[\frac{m^4\phi^2 e^{12{\alpha}}}{p_{\alpha}^2}-m^2e^{6\alpha}\right](\Delta\phi)^2,\nonumber\\
&\qquad\nonumber\\
\dot{(\Delta p_{\phi})^2}&=\frac{4m^2\phi e^{6{\alpha}}p_{\phi}}{p_{\alpha}^2}(\Delta p_{\phi})^2+4\left[-m^2e^{6\alpha}+\frac{m^4\phi^2 e^{12{\alpha}}}{p_{\alpha}^2}\right]\Delta(\phi p_{\phi}).\nn
\end{align}

As in \cite{bht1,bht2}, it is straightforward to show that the evolving degrees of freedom in the $\alpha$--\emph{Zeitgeist}, i.e.\ $\phi, p_{\phi}, (\Delta \phi)^2, (\Delta \phi p_{\phi})$ and $(\Delta p_{\phi})^2$, can be consistently chosen to be real if $\alpha$ picks up the imaginary part (\ref{imt}) (with $q_1,p_1$ replaced by $\alpha,p_\alpha$).
The set in (\ref{eom_alpha}) can be solved numerically, yielding the evolution of the transient observables of the $\alpha$--\emph{Zeitgeist} (i.e., the correlations of the evolving variables with $\Re[\alpha]$).


As generally discussed in Sec.~\ref{sec:general_transf}, the
$\alpha$--\emph{Zeitgeist} possesses only a transient validity
because $\alpha$ is a nonglobal clock. To remedy this issue in the
turning region(s) of $\alpha$, we will choose $\phi$ as the new
clock and evolve the system in the $\phi$--\emph{Zeitgeist} instead.

\subsubsection{Evolution in $\phi$}\label{sec_evphi}

The $\phi$--\emph{Zeitgeist},
\begin{align}
(\Delta\phi)^2=\Delta(\alpha\phi)=\Delta(\phi p_{\alpha})=0,
\end{align}
by solving $C_\phi$, leads to a saturation of the generalized uncertainty relation for the pair $(\phi,p_\phi)$.
The rest of the constraints are now given by
\begin{align}\label{constraints_phi}
C&={p}_{\phi}^2+(\Delta p_{\phi})^2-{p}_{\alpha}^2-(\Delta p_{\alpha})^2-e^{4\alpha}+m^2\phi^2e^{6\alpha}+(18m^2\phi^2 e^{6\alpha}-8e^{4\alpha})(\Delta\alpha)^2,\nonumber\\
C_{\alpha}&=2p_{\phi}\Delta(\alpha p_{\phi})-2p_{\alpha}\Delta(\alpha p_{\alpha})-i\hbar\, p_{\alpha}+(6m^2\phi^2 e^{6\alpha}-4e^{4\alpha})(\Delta\alpha)^2,\nonumber\\
C_{p_{\alpha}}&=2p_{\phi}\Delta(p_{\alpha}p_{\phi})-2p_{\alpha}(\Delta p_{\alpha})^2+(6m^2\phi^2 e^{6\alpha}-4e^{4\alpha})\Delta(\alpha p_{\alpha})-i\hbar\,(3m^2\phi^2 e^{6\alpha}-2e^{4\alpha}),\nonumber\\
C_{p_{\phi}}&=2p_{\phi}(\Delta p_{\phi})^2-2p_{\alpha}\Delta(p_{\alpha}p_{\phi})+(6m^2\phi^2 e^{6\alpha}-4e^{4\alpha})\Delta(\alpha p_{\phi})-2i\hbar\,m^2\phi e^{6\alpha},
\end{align}
and, again, can be used to solve for the unphysical moments $\Delta(\alpha p_{\phi}),\Delta(p_{\alpha}p_{\phi}),(\Delta p_{\phi})^2$.

The Hamiltonian constraint in the $\phi$--\emph{Zeitgeist} reads
\begin{align}\label{Cham_phi}
C_H&={p}_{\phi}^2-{p}_{\alpha}^2-e^{4\alpha}+m^2\phi^2e^{6\alpha}-\left[1-\frac{p_{\alpha}^2}{p_{\phi}^2}\right](\Delta p_{\alpha})^2-\frac{p_{\alpha}}{p_{\phi}^2}(6m^2\phi^2 e^{6\alpha}-4e^{4\alpha})\Delta(\alpha p_{\alpha})\nonumber\\
&\qquad+\left[18m^2\phi^2 e^{6\alpha}-8e^{4\alpha}+\frac{(3m^2\phi^2
e^{6\alpha}-2e^{4\alpha})^2}{p_{\phi}^2}\right](\Delta\alpha)^2+i\hbar\,\frac{m^2\phi
e^{6\alpha}}{p_{\phi}}
\end{align}
and generates the following set of equations of motion for $\alpha,p_\alpha,(\Delta\alpha)^2,(\Delta p_\alpha)^2$ and $\Delta(\alpha p_\alpha)$ which constitute the evolving degrees of freedom in the $\phi$--\emph{Zeitgeist}
\begin{align}
\dot{\phi}&=2p_{\phi}-\frac{2p_{\alpha}^2}{p_{\phi}^3}(\Delta p_{\alpha})^2+\frac{p_{\alpha}}{p_{\phi}^3}\left(12m^2\phi^2 e^{6\alpha}-8e^{4\alpha}\right)\Delta(\alpha p_{\alpha})-\frac{(6m^2\phi^2 e^{6\alpha}-4e^{4\alpha})^2}{2p_{\phi}^3}(\Delta\alpha)^2\nonumber\\&\qquad-i\hbar\,\frac{m^2\phi e^{6\alpha}}{p_{\phi}^2},\nonumber\\
&\qquad\nonumber\\
\dot{p_{\phi}}&=-2m^2\phi e^{6\alpha}+\frac{12p_{\alpha}}{p_{\phi}^2}m^2\phi e^{6\alpha}\Delta(\alpha p_{\alpha})-\left[36m^2\phi e^{6\alpha}+\frac{12m^2\phi e^{6\alpha}(3m^2\phi^2 e^{6\alpha}-2e^{4\alpha})}{p_{\phi}^2}\right](\Delta\alpha)^2\nonumber\\&\qquad-i\hbar\,\frac{m^2e^{6\alpha}}{p_{\phi}},\nonumber
\end{align}
\begin{align}
\dot{\alpha}&=-2p_{\alpha}+\frac{2p_{\alpha}}{p_{\phi}^2}(\Delta p_{\alpha})^2-\frac{6m^2\phi^2 e^{6\alpha}-4e^{4\alpha}}{p_{\phi}^2}\Delta(\alpha p_{\alpha}),\\
&\qquad\nonumber\\
\dot{p_{\alpha}}&=4e^{4\alpha}-6m^2\phi^2 e^{6\alpha}+\frac{p_{\alpha}}{p_{\phi}^2}(36m^2\phi^2 e^{6\alpha}-16e^{4\alpha})\Delta(\alpha p_{\alpha})-i\hbar\,\frac{6m^2\phi e^{6\alpha}}{p_{\phi}}\nonumber\\&\qquad-\left[108m^2\phi^2 e^{6\alpha}-32e^{4\alpha}\nonumber+\frac{(18m^2\phi^2 e^{6\alpha}-8e^{4\alpha})(6m^2\phi^2 e^{6\alpha}-4e^{4\alpha})}{p_{\phi}^2}\right](\Delta\alpha)^2,\nonumber\\
&\q\nn\\
\dot{(\Delta\alpha)^2}&=-4\left[1-\frac{p_{\alpha}^2}{p_{\phi}^2}\right]\Delta(\alpha p_{\alpha})-\frac{p_{\alpha}}{p_{\phi}^2}(12m^2\phi^2 e^{6\alpha}-8e^{4\alpha})(\Delta\alpha)^2,\nonumber\\
&\qquad\nonumber\\
\dot{\Delta(\alpha p_{\alpha})}&=-2\left[1-\frac{p_{\alpha}^2}{p_{\phi}^2}\right](\Delta p_{\alpha})^2
-2\left[18m^2\phi^2 e^{6\alpha}-8e^{4\alpha}+\frac{(3m^2\phi^2 e^{6\alpha}-2e^{4\alpha})^2}{p_{\phi}^2}\right](\Delta\alpha)^2,\nn\\&\qquad\nonumber\\
\dot{(\Delta
p_{\alpha})^2}&=\frac{p_{\alpha}}{p_{\phi}^2}(12m^2\phi^2
e^{6\alpha}-8e^{4\alpha})(\Delta
p_{\alpha})^2
-4\left[18m^2\phi^2
e^{6\alpha}-8e^{4\alpha}+\frac{(3m^2\phi^2
e^{6\alpha}-2e^{4\alpha})^2}{p_{\phi}^2}\right]\Delta(\alpha
p_{\alpha}).\nn
\end{align}

Once more, the clock variable $\phi$ develops a complex nature, in agreement with (\ref{imt}), $\Im[\phi]= -\frac{\hbar}{2p_{\phi}}$, while the evolving degrees of freedom can be chosen to be real.

\subsubsection{Numerical results}

We now wish to analyze the numerical behavior of the truncated effective
system that starts off peaked about classical trajectories for which
neither the scalar field nor the scale factor is good global
clock. For simplicity, we restrict our attention to a special class
of trajectories---those that have very few local extrema in the scale
factor: in a more general case, the internal clocks would need to be
switched many times in order to evolve through the bouncing part of
the trajectory. The cases considered are sufficient to illustrate
several qualitative points that apply more generally;
in particular, that changing the clock in the region of maximal expansion will not work in a generic solution.
\begin{center}
\begin{figure}[htbp!]
\psfrag{h}{$\hbar$}
 \subfigure[]{\label{fig:1}\includegraphics[width=7.4cm]{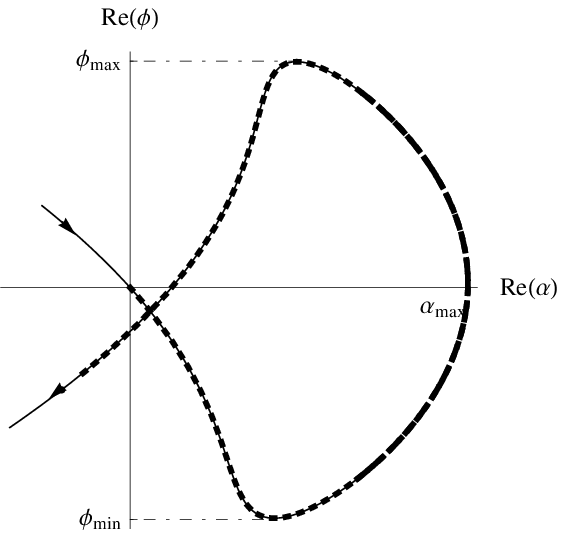}}
 \hspace{.5cm}   \subfigure[]{\label{fig:2a}\includegraphics[width=7.4cm]{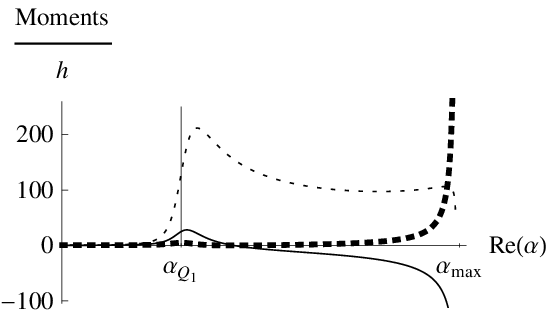}}\\
  \caption{\small (a) Classical trajectory (dotted) and patched up effective trajectory: $\alpha$--gauge (solid), $\phi$--gauge (dashed). (b) Moments in $\alpha$--gauge on the incoming branch: $(\Delta \phi)^2$ (thick, dashed), $(\Delta p_{\phi})^2$ (thin, dashed), $\Delta(\phi p_{\phi})$ (solid). $\alpha_{Q_1}$ is the quasi--turning point of $\alpha$ on the incoming branch, where the clock becomes `slow' (see discussion and Fig.~\ref{fig:3a}).}
\end{figure}
\end{center}
\begin{center}
\begin{figure}[htbp!]
\psfrag{h}{$\hbar$}
 \subfigure[]{\label{fig:2b}\includegraphics[width=7.4cm]{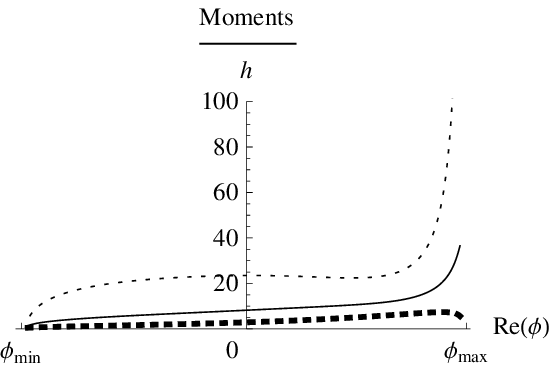}}
    \hspace{.5cm}\subfigure[]{\label{fig:2c}\includegraphics[width=7.4cm]{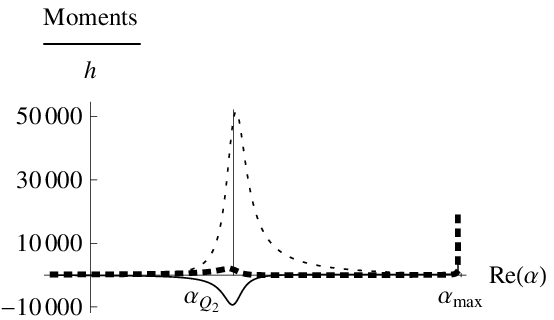}}\\
  \caption{\small (a) Moments in $\phi$--gauge: $(\Delta \alpha)^2$
(thick, dashed), $(\Delta p_{\alpha})^2$ (thin, dashed),
$\Delta(\alpha p_{\alpha})$ (solid). (b) Moments in $\alpha$--gauge
on the outgoing branch: $(\Delta \phi)^2$ (thick, dashed), $(\Delta
p_{\phi})^2$ (thin, dashed), $\Delta(\phi p_{\phi})$ (solid).
$\alpha_{Q_2}$ is the quasi--turning point of $\alpha$ on the
outgoing branch, where the clock becomes `slow' (see discussion and
Fig.~\ref{fig:3b}).}
\end{figure}
\end{center}
\begin{center}
\begin{figure}[htbp!]
 \subfigure[]{\label{fig:3a}\includegraphics[width=7.4cm]{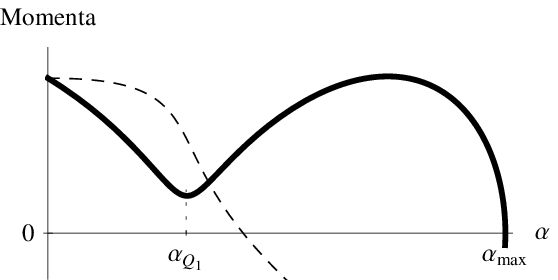}}
    \hspace{.3cm}\subfigure[]{\label{fig:3b}\includegraphics[width=7.4cm]{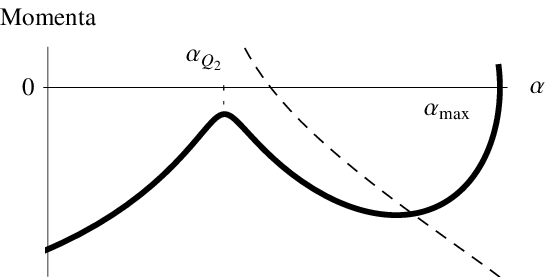}}\\
  \caption{\small (a) Classical momenta on the incoming branch with quasi--turning point $\alpha_{Q_1}$ of the clock $\alpha$:
$p_{\phi}$ (dashed), $p_{\alpha}$ (solid). (b) Classical momenta on the outgoing branch with quasi--turning point $\alpha_{Q_2}$:
$p_{\phi}$ (dashed), $p_{\alpha}$ (solid). }
\end{figure}
\end{center}
\vspace{-2.99cm}

Figure~\ref{fig:1} displays an effective
relational
trajectory in the
configuration space that was patched together by first evolving it
using $\alpha$\ as a clock, followed by transforming to the $\phi$--\emph{Zeitgeist} between the extremal points $\phi=\phi_{min}$\ and
$\alpha=\alpha_{max}$, finally switching back to the $\alpha$--\emph{Zeitgeist} after $\alpha=\alpha_{max}$, but before $\phi=\phi_{max}$. We have switched gauges and clocks by the general method developed in Sec.~\ref{sec:general_transf}, but abstain from explicitly exhibiting the corresponding formulae. Alongside the effective trajectory,
Fig.~\ref{fig:1} displays the corresponding classical trajectory,
with the two being virtually indistinguishable. For the particular
numerical evolution plotted, we chose the quantum scale such that
$\sqrt{\hbar}\sim10^{-4}$\ when compared to the expectation values
that are of order $1$. The leading--order quantum corrections are of
order $\hbar$\ and are therefore $\sim10^{-8}$\ times weaker than
the classical effects. In this regime the quantum backreaction is
virtually nonexistent, and the classical variables evolve
essentially independently from the quantum modes. The necessity for
this large separation of the classical and quantum scales chosen
ultimately traces back to the classical chaoticity of the system and
can be illustrated by the behavior of the moments in
Figs.~\ref{fig:2a}--\ref{fig:2c}. The initial values of the moments
in the $\alpha$--\emph{Zeitgeist} are close to $\hbar$; however, at
a certain point in the outgoing trajectory, they are about $10^4$\
times larger than their initial values, which makes the assumption
about the semiclassical falloff outright inapplicable if the
separation of the different perturbative orders is less than $10^4$.
The defocusing of classical trajectories in the region of maximal
expansion forces a semiclassical state initially peaked on nearby
classical trajectories to inevitably spread apart yielding an
overall growth of the moments. For the classical solution reproduced
by the effective solution in Fig.~\ref{fig:1}, the defocussing of
initially neighboring trajectories is displayed in
Fig.~\ref{fig:traj_chaos}.

Furthermore, the ``spikes'' in the moments, particularly $(\Delta
p_{\phi})^2$, in Figs.~\ref{fig:2a} and~\ref{fig:2c} trace their
origin to the classical quasi--turning points of the internal clock
$\alpha=\alpha_{Q_1}$\ and $\alpha=\alpha_{Q_2}$, where
$\dot{\alpha}=-2p_{\alpha}$\ is small and the clock $\alpha$ thus
becomes `too slow' for resolving the evolution of other degrees of
freedom with respect to it (also see the discussion in
Sec.~\ref{sec_failzeit}). One might suggest evolving
through these regions using $\phi$\ as the internal clock; however,
this may not be feasible in general, as the quasi--turning points in $\alpha$\
may lie too close to the turning points in $\phi$: for the
particular trajectory, this is illustrated in Figs.~\ref{fig:3a}
and~\ref{fig:3b} for the incoming and outgoing branches,
respectively, where one can see the proximity of the local minima in
$p_{\alpha}$\ and the points where $p_{\phi}=0$. It can be concluded from the general characterization of classical solutions to this model given in \cite{kkt} that this property is a generic one in the space of solutions.
Both $\alpha$ {\it and} $\phi$ (as well as their momenta which also feature turning points) are thus `poor clocks' for the same piece of the trajectory, leading to a poor resolution of relational evolution and thus to a large growth of the moments. But if $\alpha$ and $\phi$ fail to be good clocks, neither could any functions $f(\alpha)$ or $g(\phi)$ serve as better clocks for such a trajectory in this region because $(\Delta f)^2\propto(\Delta\alpha)^2$ and $(\Delta g)^2\propto(\Delta\phi)^2$, and since both $(\Delta\alpha)^2,(\Delta\phi)^2$ cannot consistently vanish in this region (see section \ref{sec_failzeit}), no $f$-- or $g$--Zeitgeist could be valid either. It is difficult to {\it explicitly} demonstrate from this alone that {\it no} other phase space function could serve as a better clock in such a region. Nevertheless, while for the particular trajectory exhibited here the accompanying large growth of moments is still within the validity of the semiclassical truncation, we shall argue shortly on more general grounds, that for more generic trajectories this will become a fundamental problem that prevents clock changes and relational evolution altogether.
\begin{center}
\begin{figure}[htbp!]$\begin{array}{cc}
\includegraphics[width=7.4cm]{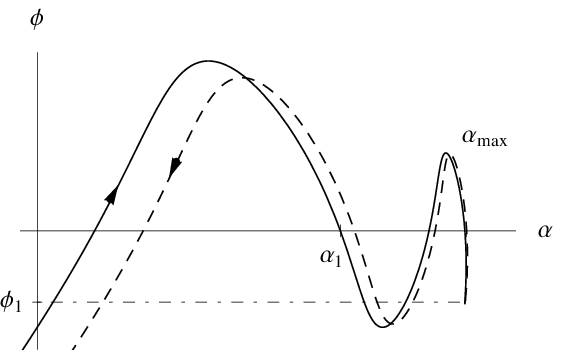} &
\includegraphics[width=7.4cm]{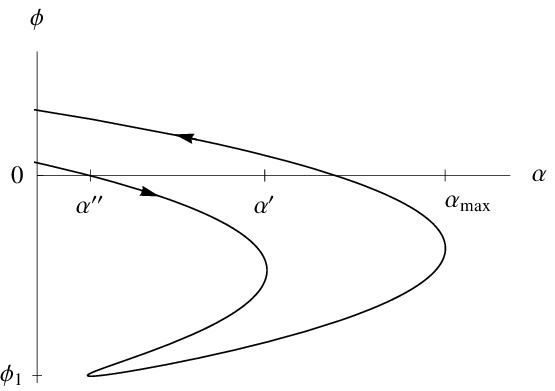} \end{array}$
\caption{\small \label{fig:4} Left: a classical configuration space
trajectory computed using the same model parameters as in
Fig.~\ref{fig:1}, but with different initial conditions: incoming
branch (solid), outgoing branch (dashed). Right: a closeup of the
same trajectory near $\alpha=\alpha_{max}$; there are two other
local extrema in $\alpha$\ labeled by $\alpha'$ (a maximum) and
$\alpha''$ (a minimum), in addition, there $\phi$ reaches a locally
minimal value $\phi_1$ very near $\alpha=\alpha''$.}
\end{figure}
\end{center}
\begin{center}
\begin{figure}[htbp!]
\psfrag{h}{$\hbar$}
\includegraphics[width=7.4cm]{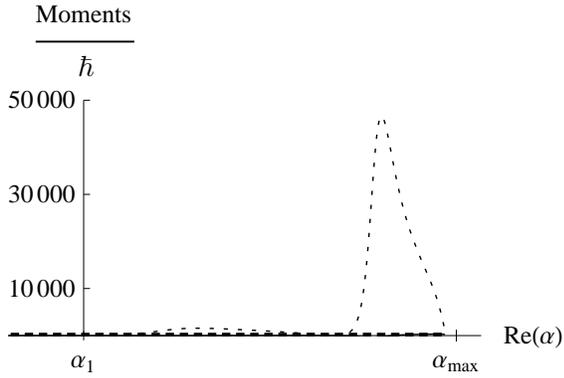}
\caption{\small \label{fig:5} Moments in $\alpha$--gauge on the
incoming branch evolved effectively in a state initially peaked
around the trajectory in Fig.~\ref{fig:4} : $(\Delta \phi)^2$
(thick, dashed), $(\Delta p_{\phi})^2$ (thin, dashed), $\Delta(\phi
p_{\phi})$ (solid).}
\end{figure}
\end{center}
\vspace{-2cm}

The above problem is a manifestation of a more general issue in this
model: an arbitrary classical trajectory can exhibit structure, such
as local maxima and minima, at all scales---there is no natural
threshold scale below which there is no classical structure. We
illustrate this further by picking a slightly more complicated
classical trajectory, plotted in Fig.~\ref{fig:4}. This trajectory
uses the same model parameters (namely, $m$ and $\hbar$) as the one
in Fig.~\ref{fig:1}, with different initial conditions. The scale
factor exhibits not one, but three local extrema at
$\alpha=\alpha_{max}, \alpha', \alpha''$. The corresponding
effective system is much more unstable already along the incoming
branch when evolved in $\alpha$, so much so that by the time it
approaches the classical turning points in $\alpha$, the spreads are
of order comparable to the separation between the three extrema of
$\alpha$ (Fig.~\ref{fig:5}). The situation is, in fact, even worse,
as the separation between the turning point of $\phi$, where
$\phi=\phi_1$, and the local minimum in $\alpha$, where
$\alpha=\alpha''$, is of order $\sqrt{\hbar}$, and the two points
could not be resolved even if the moments remained well behaved; no clock change between $\alpha$ and $\phi$ can be performed.
Therefore, given the chosen quantum scale, this fairly benign trajectory cannot be
resolved by the effective evolution as it stands. For more generic classical trajectories, such as those displayed in Fig.~\ref{fig:class_sol}, the problem must magnify: as can be inferred from the general discussion in \cite{kkt}, for any given choice of the quantum scale there will
be an infinite set of classical trajectories with extrema in $\alpha$\ and $\phi$\
separated on or below that scale.

\subsubsection{Breakdown of relational evolution}

Let us now argue that for generic semiclassical trajectories this wealth of structure on all scales will impede clock changes and, indeed, ultimately lead to a generic breakdown of quantum relational evolution in the region of maximal expansion. This can already be deduced from the classical dynamics. Take an arbitrary classical phase space trajectory. Pick an arbitrary open neighborhood of the constraint surface through which this orbit passes (in fact, it can be any open neighborhood because a generic trajectory passes through any open neighborhood of the constraint surface in chaotic systems.) Any phase space function that is to serve locally as a good relational clock for this orbit in this open neighborhood must grow monotonically along the trajectory (ideally, one would like the level surfaces of this clock function to be orthogonal to the tangent vector of the trajectory at each point). That is to say, this phase space function must vary on the same scales as the trajectory itself. As a consequence of the instability of initial data in this chaotic model, a trajectory which initially is arbitrarily close to the particular trajectory we are considering will generically experience a completely uncorrelated evolution. Therefore, in our open neighborhood, the phase space function that served as a good clock for the first trajectory will generically fail to be a good clock for the second trajectory and thus is highly orbit dependent. Indeed, pick an arbitrarily small neighborhood on the constraint surface. The system being chaotic implies that infinitely many trajectories will pass through this neighborhood in uncorrelated manner and in all directions. Since any clock function at every phase space point defines an internal time direction which is orthogonal to its level surfaces, it is evident that no phase space function can be a good clock for all trajectories in the entire neighborhood (recall that generic trajectories can vary on arbitrary scales). 

This has severe repercussions for the effective semiclassical trajectories. On account of the moments and spreads, at semiclassical order $\hbar$, evolving an effective solution through the quantum phase space means evolving a neighborhood of volume of the order $\hbar^2$ through the phase space (it is of order $\hbar^2$ because we have four canonical variables). This can work for initially highly semiclassical states corresponding to superpositions of special classical solutions featuring very few turning points and varying roughly on the same scales, as seen in the example above. However, for an initially semiclassical effective state corresponding to a superposition of generic classical orbits that follow unrelated cosmological fates, a relational evolution must break down in the region in which these classical trajectories scatter apart. In fact, in this case there may exist neighborhoods in which a phase space function could serve as a good clock for a given classical trajectory contained in the semiclassical superposition. But as noted above, such a clock would be highly orbit dependent, and so in the generic case no neighborhood with volume of order $\hbar^2$ (or larger) exists in such a scattering region for which {\it any} phase space function could be a useful clock for {\it all} classical orbits contained in the superposition. But this would be required in order to evolve a semiclassical state relationally. At this stage, no clock change is possible, and relational evolution must break down altogether. For such trajectories it is then
fundamentally impossible, using the effective method, to construct entire semiclassical states which evolve nicely through the region of maximal expansion. Even more, effective relational evolution (in a `classical' clock) must break down for general (nonsemiclassical) effective states: if one attempted to evolve a neighborhood larger than of the order $\hbar^2$ through the quantum phase space, the above mentioned problems can only intensify.

\section{Conclusions}\label{sec_con}

The present work is a first step in the study of relational quantum
dynamics in the generic nonintegrable case featuring a
nontrivial coupling between the clock and the evolving degrees of
freedom. The effective approach \cite{bht1,bht2} seems especially
well geared for investigating semiclassical relational dynamics of
nonintegrable systems because it enables one to make sense of
temporally local time evolution, yielding transient relational
observables and allowing one to switch back and forth between
various clock variables.

In particular, we have applied the effective approach to the (nonintegrable) closed FRW model universe filled with a minimally coupled massive scalar field whose quantum dynamics has thus far not been properly studied. The numerical results obtained here for rather benign trajectories already demonstrate that semiclassicality in this cosmological model is a delicate issue in the region of maximal expansion and generally fails due to the sensitivity of solutions to the initial conditions which results in a generic defocusing of classical trajectories in this region; a semiclassical state peaked on initially nearby classical trajectories inevitably has to spread apart. This distinguishes the present cosmological model from the toy models earlier studied in \cite{bht2}, where coherent states are available which are sharply peaked even in the turning region of the nonglobal clocks which, furthermore, are decoupled such that the `imperfect' behavior of one clock does not depend on that of the other.

The region of maximal expansion, in fact, features a chaotic scattering \cite{cornshell}, which renders it especially challenging for relational dynamics. Indeed, the effective results reported here provide evidence that allows us to argue that relational dynamics, while possible for sufficiently sharply peaked states, generically breaks down in the region of maximal expansion. In this regime, we can no longer trust the effective semiclassical truncation, since the moments grow beyond order $\sqrt{\hbar}$ and eventually diverge despite quantum backreaction not playing a prominent role. A generic classical trajectory exhibits quasi--turning points of the clock $\alpha$ immediately following/preceding a turning point of the field $\phi$ \cite{kkt}; the two clock momenta thus become small (or vanish) in the immediate neighborhood of each other, rendering both clocks `too slow' to properly resolve relational evolution \cite{bht1,bht2,marolf1,marolf2,gidmarhar,biajoh1} and yielding large uncertainties. (The momenta of the two clocks do not fare any better as clocks themselves because they are generically highly oscillatory in nature.) We have argued on general grounds that, in the generic case, no change of clock and \emph{Zeitgeist} can remedy this, and that in the region of maximal expansion no good clock function can exist for effective dynamics on account of the wealth of structure on all scales present in chaotic models. In agreement with the general discussion in \cite{bht1}, the failure of the effective semiclassical truncation in this manner is strong evidence, suggesting that relational evolution generally breaks down due to a mixing of internal time directions in such a regime. This is the effective analogue of nonunitarity in a (local) deparametrization at a Hilbert space level resulting in a breakdown of (any) inner product based on level surfaces of the nonglobal clock variable.

The generic breakdown of semiclassicality in the region of maximal expansion is compatible with the necessary breakdown of the WKB approximation to (\ref{wdw1}) earlier reported in the literature \cite{page2,Kiefer,cornshell,lafshell,zehbook}. Note, however, that while the WKB approximation is a specific method to study the semiclassical limit, here we have employed a very general semiclassical approximation. Indeed, in contrast to the arguments put forward in \cite{zeh,Kiefer,zehbook} concerning the semiclassical limit as obtained by WKB techniques, in the effective approach it {\it is} possible to obtain semiclassical solutions which follow a classical trajectory {\it without} continuous measurement through higher degrees of freedom if the state is initially sufficiently sharply peaked and the corresponding classical trajectory sufficiently benign. One merely has to switch the relational clock at intermediate stages according to the general construction presented in this article. For these sufficiently peaked states, the (relational) IVP retains its classical (deterministic) character of having the recollapsing branch as the logical successor of the expanding one---in contrast to the discussion in \cite{zeh,Kiefer,zehbook}---although, clearly, the recovery of a `good (temporally) local relational evolution' depends very sensitively on the state.

As a consequence of nonintegrability being the generic case in
dynamical systems \cite{chaos}, we conjecture that the reported
qualitative results concerning (the breakdown of) relational
evolution and semiclassicality (ultimately rooted in the
nonintegrability) should feature prominently in a generic
situation in quantum cosmology and gravity, thus emphasizing the
delicate nature of the question posed in the beginning. Generically,
`good relational evolution' appears to be only a transient and
semiclassical phenomenon.

However, nonintegrability manifests itself in sensitivity to
initial data. A natural question to ask is whether
the quantum-modified dynamics of
loop quantum cosmology,
which originates from
the minimal area gap \cite{martinbooks,lqc} (giving
`infinitesimally close' a different meaning), could possibly resolve
some of the chaotic attributes of this model universe, thereby
providing a `better behaved' theory.

\section*{Acknowledgements}

The authors would like to thank Martin Bojowald for valuable incentives and suggestions, and Renate Loll and David Sloan for interesting discussions.
E.K.\ is supported by the Czech government grant agency under Contract No.\ GA \v{C}R $202/08/$H$072$. Furthermore, E.K.\ would like to express gratitude to the Institute for Gravitation and the Cosmos at Pennsylvania State University for hospitality during the first stages of this work.

\end{document}